\documentclass[
pre, %preprint,
twocolumn,
superscriptaddress, a4paper, showpacs,
]{revtex4}
\usepackage{graphicx}
\usepackage{amsmath,amssymb}

\makeglossary

\begin{document}

%\title{Oscillatory zoning in ternary solid solutions with weak nonideality \\
%    grown from aqueous solution with different species solubilities}
%\title{Growth instability of solid solutions caused by system asymmetry}
\title{Different routes towards oscillatory zoning in the growth of solid solutions}

\date{\today}

\author{Ihor Lubashevsky}
 \email{ialub@fpl.gpi.ru}
 \affiliation{Westf\"alische Wilhelms Universit\"at
M\"unster, Institut f\"ur physikalische Chemie, Corrensstr. 30, 48149
M\"unster, Germany}
 \affiliation{A.M. Prokhorov General Physics Institute, Russian Academy of
Sciences, Vavilov Str. 38, Moscow, 119991 Russia}
\affiliation{Moscow Technical University of Radioengineering, Electronics, and Automation,
    Vernadsky 78, 119454, Moscow Russia}
\author{Tanja Mues}
 \email{t_mues01@uni-muenster.de}
 \affiliation{Westf\"alische Wilhelms
Universit\"at M\"unster, Institut f\"ur physikalische Chemie, Corrensstr. 30,
48149 M\"unster, Germany}
 \affiliation{Center of Nonlinear Science CeNoS,
Westf\"alische Wilhelms Universit\"at M\"unster, 48149 M\"unster, Germany}
\author{Andreas Heuer}
 \email{andheuer@uni-muenster.de}
 \affiliation{Westf\"alische Wilhelms
Universit\"at M\"unster, Institut f\"ur physikalische Chemie, Corrensstr. 30,
48149 M\"unster, Germany}
 \affiliation{Center of Nonlinear Science CeNoS,
Westf\"alische Wilhelms Universit\"at M\"unster, 48149 M\"unster, Germany}

\begin{abstract}
Oscillatory zoning, i.e. self-formation of spatial quasi-periodic oscillations in the composition of solid growing from aqueous solution, is analyzed theoretically. Keeping in mind systems like (Ba,Sr)SO$_4$ we propose a 1D model that takes into account the nonideality of the solid solution and the system asymmetry, in particular, reflecting itself in different solubilities for such systems. Based on a linear stability analysis different parameter regions can be identified. Even an ideal solution solution with a sufficiently large asymmetry can display oscillatory zoning.  Numerical simulations complement the linear stability analysis as well as the qualitative consideration of the instability development and reveal the nature of the limit cycles.

\end{abstract}


\maketitle

\section{Introduction}

Oscillatory zoning (OZ), i.e. spatial pattern made of
quasiperiodic variations of the solid composition from core of
crystals to their rim is widely met in natural  minerals (see,
e.g., Ref.~\cite{ShoreFowler}). The appearance of such patterns
was traditionally related to cyclic changes in surroundings during
crystal formation in rocks. However the success of reproducing OZ
in calcite crystals \cite{Reeder} and (Ba,Sr)SO$_4$ solid
solutions \cite{Putn1,Putn2,Putn3} in laboratory under
quasistationary conditions has demonstrated the fact that at least
partly OZ can result from selforganization during crystal growth
in solution.

\begin{figure}[h]
\begin{center}
\includegraphics[width=0.75\linewidth]{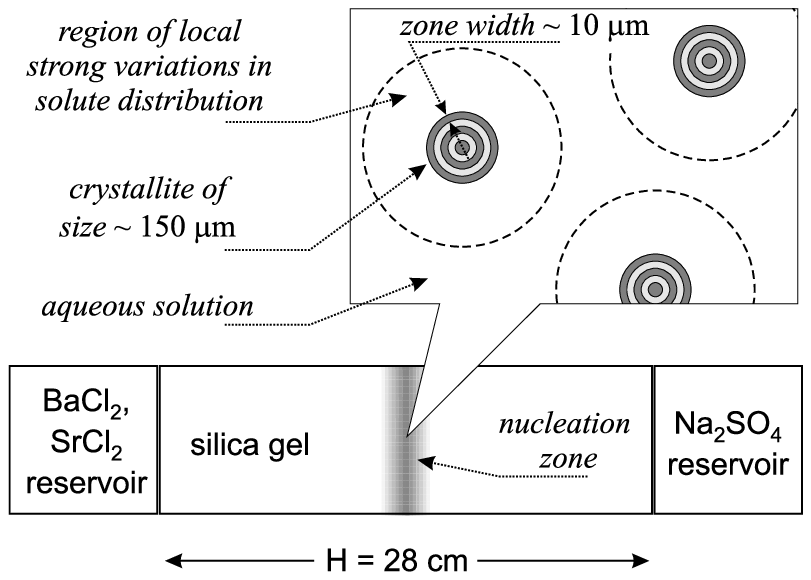}
\end{center}
\caption{\label{Fig.1} Experimental setup in which oscillatory zoned crystals
of (Ba,Sr)SO$_4$ were synthesized by Putnis \textit{et
al.}~\cite{Putn1,Putn2,Putn3}. The reactants counterdiffuse in the column and
(Ba,Sr)SO$_4$ crystals nucleate. The upper window sketches the structure of the
nucleation zone and the length scales involved.}
\end{figure}

The experimental  setup used by Putnis \textit{et al.}
\cite{Putn1,Putn2,Putn3} is sketched in Fig.~\ref{Fig.1}. It
consists of two reservoirs, one filled with aqueous solution of
BaCl$_2$/SrCl$_2$ and the other with Na$_2$SO$_4$. The two
reservoirs are connected by a column filled with silica gel to
inhibit convective transport. At the beginning of experiments the
reactants start to diffuse toward each other through the column.
As the diffusion fields of Ba$^{2+}$, Sr$^{2+}$, and SO$_4{}^{2-}$
overlap and the solute concentration product exceeds the
nucleation threshold in the vicinity of the column center, the
crystal nuclei form. In approximately one month the experiments
were terminated. The obtained crystals exhibited OZ although no
external fluctuations were imposed on the system.

Following the spirit of the general model by Ortoleva
\cite{Ortoleva1,Ortoleva2} for the growth instability caused by
autocatalytic interaction of the species at the crystal surface
L'Heureux \textit{et al.} \cite{LH1,LH2,LH3} proposed a rather
sophisticated model for OZ in the (Ba,Sr)SO$_4$ solid growing from
aqueous solution. The detailed analysis of these models was
carried out within the boundary layer approximation.

In a previous paper \cite{we} we have demonstrated that OZ in
crystals growing from solution can be described as a
boundary-reaction-diffusion problem. It is characterized by
passive diffusion of species through the solution bulk to the
crystal surface where their interaction gives rise to the crystal
growth. The latter, however, proceeds with a very low rate so that
the crystal boundary can be treated as a surface fixed in space.
It that work we have mainly studied the presence of the
instability with respect to the nonideality parameter $\theta$. It
turned out that for sufficiently large $\theta$, i.e. $\theta
> \theta_{c0}$ the instability and thus OZ can indeed be observed.

Experimentally, however, it is observed that in particular solid solutions which very different solubility products of the endmembers display OZ (such as (Ba,Sr)SO$_4$) whereas systems with similar solubility products (such as (Ba,Sr)CO$_3$) do not display OZ \cite{Prieto97}. For example, for the first case the solubility product of both endmembers differs by three orders of magnitude (see, e.g., \cite{Therm}). The solubility product is related to the system asymmetry $\phi$. Thus, the question emerges whether the model also allows OZ for systems with a pronounced asymmetry rather than a significant nonideality.

The purpose of this work is fourfold. First, we rederive our model in a somewhat extended way which allows one to better understand the microscopic origin of the different parameters. The definitions have been chosen such that the final model equations are identical to the model studied in our previous work \cite{we}. Second, after deriving the somewhat complex instability conditions from the linear stability analysis we argue on a semi-quantitative level that indeed the model possesses an additional instability channel for sufficient large values of the system asymmetry and thus obtain a semi-quantitative phase diagram of the instability region. Third, via a careful mathematical analysis we somewhat modify this picture, yielding some surprising features in the newly analyzed instability regime. Fourth, via numerical simulations we illustrate the behavior beyond the linear regime.

\section{Model \label{sec:2}}

\subsection{Energetics of crystal growth}

We take into account the following mechanism of crystal growth
having in mind the (Ba,Sr)SO$_4$. The ions SO$_{4}^{2-}$ (below
species of type 0), Ba$^{2+}$ (species 1), and Sr$^{2+}$ (species
2) diffuse to the crystal surface through the aqueous solution,
where they are adsorbed and display surface diffusion. If they
reach the atomic steps they are incorporated into the crystalline
lattice via the following precipitation reactions
\begin{subequations}\label{Intr:5}
\begin{align}
    \label{Intr:5a}
    \text{Ba}^{2+} + \text{SO}_4{}^{2-} &\rightarrow \text{BaSO}_4 & &\text{(channel 0--1)}\,,\\
    \label{Intr:5b}
    \text{Sr}^{2+} + \text{SO}_4{}^{2-} &\rightarrow \text{SrSO}_4 & &\text{(channel 0--2)}
\end{align}
\end{subequations}
 The latter process is considered
to be irreversible, i.e. the solid dissolution is ignored, which
means the system to be far from thermal equilibrium and the growth
rate cannot take too low values in the case under consideration.
Finally they are forming a new layer of the crystal.

Migrating along the crystal surface adatoms experience many
different local environments depending on the surface composition
which will be characterized by the mole fraction $\chi$ of species
1 ($0\leq\chi\leq1$). The competition between adsorption and
desorption is determined by the effective adsorption energies
$E_i(\chi)$ ($i = 0$, 1, 2) reflecting the species interaction
with the crystal surface and aqueous solvent. In the mean field
approximation they are written as
\begin{subequations}\label{aden}
\begin{align}
E_{0}(\chi)  &  =\epsilon_{0}-g_{1}\chi-g_{2}(1-\chi)\,,\label{aden0}\\
E_{1}(\chi)  &  =\epsilon_{1}-g_{1}+\theta(1-\chi)\,,\label{aden1}\\
E_{2}(\chi)  &  =\epsilon_{2}-g_{2}+\theta\chi\,,\label{aden2}
\end{align}
\end{subequations}
where all the energy quantities are measured in units of
temperature, $\epsilon_i$ is the solvation energy of species $i$,
the constant $g_i$ characterizes the interaction between adatoms
of type $i = 1$, 2 with atoms of type 0 lying in the surface
atomic layer of the crystal lattice, and the parameter $\theta>0$
quantifies the solid solution nonideality. This expresses the fact
that the strongest interaction on the crystal surface holds
between like ions.

In these terms equilibrium between the adsorbed layer and the
aqueous solution region adjacent to the crystal surface implies
the following relation between the adatom concentrations $c_i$ and
the concentration $C^s_i$ of the corresponding species near the
crystal surface
\begin{equation}\label{tau1}
c_i = a C_i^s e^{-E_{i}(\chi)}\,,
\end{equation}
where $a$ is the characteristic size of the crystalline cell.

In principle, on vicinal crystal surfaces the adatoms should have
some solvent shells and for them to be incorporated into the
crystal lattice these shells have to be destroyed. If it is
essential then the precipitation reactions~\eqref{Intr:5} at the
surface atomic steps limit the crystal growth and the adsorbed
layer can be assumed to be in quasiequilibrium, meaning
equalities~\eqref{tau1} to hold. In this case the partial rates
$\vartheta_1$ and $\vartheta_2$ of the crystal growth though
channels~\eqref{Intr:5a} and \eqref{Intr:5b}, respectively, are
given by the expression
\begin{equation}\label{pgr}
    \vartheta_i = \nu_i \frac{a^6}{l}c_0c_i\,,
\end{equation}
where $\nu_i$ is the rate at which the pair of the Ba$^{2+}$,
SO$_{4}^{2-}$ adatoms or the Sr$^{2+}$, SO$_{4}^{2-}$ adatoms
meeting at the surface steps are  incorporated in the crystal
lattice and $l$ is the mean distance between these steps.

Combining expressions~\eqref{tau1} and \eqref{pgr} we get the
desired relationship between the partial growth rates via the
channels 0--1, 0--2 and the corresponding values of the solute
concentrations $C^s_0$, $C^s_1$, and $C^s_2$ near the crystal
surface
\begin{subequations}\label{pgr1}
\begin{align}
\label{pgr1a}
    \vartheta_1 &= \varpi \left( \frac{\nu_1}{\nu_2}\right)^{1/2} e^{-\frac12
    \eta}\, e^{\phi\chi - \theta(1-\chi)}\,
    C_0^s C_1^s \,, \\
\label{pgr1b}
    \vartheta_2 &= \varpi \left( \frac{\nu_2}{\nu_1}\right)^{1/2} e^{\frac12
    \eta}\, e^{-\phi(1-\chi) - \theta\chi}\,
    C_0^s C_2^s \,.
\end{align}
\end{subequations}
Here we have introduced the kinetic coefficient
\begin{equation}\label{pgr2}
    \varpi = \sqrt{\nu_1\nu_2}\, \frac{a^8}{l} e^{2g_{12} -\epsilon_0 - \epsilon_{12}}
\end{equation}
and rewritten the interaction constants $g_{1,2}$, $\epsilon_{1,2}$ using combination of the quantities
\begin{align}
\label{pgr3}
    g_{12} & = \frac12\left(g_1+g_2\right)\,,& \phi & = g_1-g_2\,,
\\
\label{pgr4}
    \epsilon_{12} & = \frac12\left(\epsilon_1+\epsilon_2\right)\,,& \eta & = \epsilon_1-\epsilon_2
\end{align}
to mark out the difference of species 1 and 2 in properties.

Expressions~\eqref{pgr1} are actually the main result  of this
subsection and form the basis of the model for the crystal growth
to be constructed in the next section. It is rather similar to the
model we have developed previously~\cite{we}, enabling us to
sketch out the principle aspects only. Below we will assume the
inequality $\phi>0$ to hold beforehand because, otherwise, the
indices could be just exchanged.

\subsection{Model equations}

In the aqueous solution the SO$_{4}^{2-}$ ions are assumed to be
abundant. Thus, we can regard their concentration as a fixed value
$C_{0}$. In this case the crystal growth in the 1D description is
governed by the boundary-reaction-diffusion model developed in our
previous work~\cite{we}. Namely, diffusion of the components
$i=1,2$ through the solution is considered within the region $z
\in [0,L]$ and is described by the equation
\begin{equation}\label{Diffusion}
\frac{\partial C_i(z,t)}{\partial t} = D_i\frac{\partial^2 C_i(z,t)}{\partial
z^2}\,,
\end{equation}
where $D_i$ is the diffusivity of the species $i$ in the aqueous solution and the system size $L$ should be chosen
large enough in order to enable us to fix the influx of both the components at the external boundary $z=L$
\begin{equation}\label{Influx}
G_i = D_i\left. \frac{\partial C_i(z,t)}{\partial z} \right|_{z=L}.
\end{equation}
Then having in mind expressions~\eqref{pgr1} we write the following boundary condition at the crystal surface ($z=0$)
\begin{equation}\label{boundaryfin}
    D_i\left. \frac{\partial C_i(z,t)}{\partial z}
    \right|_{z=0} = \frac{aC^\mathrm{s}_i}{\tau_i(\chi)}
\end{equation}
which relates the boundary values of diffusion flux and the rates of species attachment to the crystal surface
\begin{equation}\label{attrate}
    r_i := \frac{aC^\mathrm{s}_i}{\tau_i(\chi)}
\end{equation}
caused by the growth process. Here the time scales of the crystal growth dynamics via the channels 0-1 and 0-2
individually are specified as
\begin{subequations}\label{tau}
\begin{align}
\label{tau:a}
     \tau_1(\chi) & = \tau_g \left(\frac{\nu_2}{\nu_1}\right)^{1/2}
     e^{\frac{1}{2}\eta -\phi\chi +\theta(1-\chi)}\,,
\\
\label{tau:b}
    \tau_2(\chi) & = \tau_g \left(\frac{\nu_1}{\nu_2}\right)^{1/2}
     e^{-\frac{1}{2}\eta +\phi(1-\chi) +\theta\chi}\,,
\end{align}
\end{subequations}
where the time scale of the crystal growth dynamics as a whole process is
\begin{equation}\label{tau:c}
    \tau_{g} = \frac{a^4}{\varpi C_0}\,.
\end{equation}
Finally, the solid composition is governed by the equation
\begin{equation}\label{chidot}
\frac{d\chi}{dt}=a^2\,\Bigl[\left(1-\chi\right)\,\frac{a
C^s_1}{\tau_1(\chi)}-\chi\,\frac{a C^s_2}{\tau_2(\chi)}\Bigr]
\end{equation}
following from mass conservation and used previously in a number
of papers on OZ, see, e.g., Refs.~\cite{LH1,LH2,LH3,we}.

The given system admits only one steady state solution
\begin{subequations}\label{stst}
\begin{equation}\label{stst:1}
\begin{split}
    C_{1}(z) & = C_{1,\text{st}}^{s}+\chi_\text{st} \frac{G}{D_{1}}z\,,
 \\
    C_{2}(z) &= C_{2,\text{st}}^{s}+(1-\chi_\text{st})\frac{G}{D_{2}}z\,,
\end{split}
\end{equation}
where $G = G_1+G_2$ is the total diffusion flux determining the growth rate of the crystal as a whole, the
corresponding value of the crystal composition $\chi_\text{st} = G_1/G$, so
\begin{align}\label{stst:3}
    G_1  &= \chi_\text{st}G \,,& G_2 &= (1-\chi_\text{st})G\,,
\end{align}
and by virtue of \eqref{boundaryfin} the boundary values of the species concentrations are
\begin{equation}\label{stst:4}
\begin{split}
    C^s_{1,\text{st}} & = \frac{\tau_1(\chi_\text{st})}a\chi_\text{st}G\,,
 \\
    C^s_{2,\text{st}} & = \frac{\tau_2(\chi_\text{st})}a(1-\chi_\text{st})G\,.
\end{split}
\end{equation}
\end{subequations}
It should be noted beforehand that the given model describing, generally speaking, surface kinetics contains at least
three variables, the solid state composition $\chi$ and two boundary values of the species concentrations $C_1^s$,
$C^s_2$. So the system instability can be described using the classical notions of relaxation oscillations on a
two-dimensional phase plane only at a rough approximation, as it has been already shown in our previous paper
\cite{we}.

\section{The instability domain}

\subsection{The eigenvalue problem}

Now let us analyze in a rigorous way the linear stability of the system around the steady state described by
expressions~\eqref{stst}. For this purpose the dynamics of small perturbations
\begin{equation}
    \delta C_{i}(t,z)\propto \exp \left\{ \gamma t-p_{i}z\right\} \,,\quad
    \delta \chi (t)\propto \exp \left\{ \gamma t\right\}  \label{ib:1}
\end{equation}
in the species distribution and the composition of the crystal surface is considered. Here $\gamma $ is the instability increment and the parameters $\{p_i\}$ such that $\operatorname{Re}p_{i}>0$ characterize localization of the perturbations $\delta C_{i}(t,z)$ near the crystal surface. Then the governing equations~(\ref{Diffusion})--(\ref{boundaryfin}), and (\ref{chidot}) are linearized with respect to perturbations~(\ref{ib:1}) in the vicinity of the stationary solution~(\ref{stst}). The system of algebraic equations obtained in this way gives us the eigenvalue equation for the instability increment $\gamma $. This procedure is practically identical to that from Ref.~\cite{we}. So here we skip the corresponding mathematical manipulations and write directly the desired eigenvalue equation in the final form
\begin{widetext}
\begin{equation}\label{ib:2}
    \frac{\zeta ^{2}}{g}e^{i2\psi }=-1+\chi (1-\chi )
    \left[ (\theta +\phi )\frac{
    (\zeta \Delta )e^{i\psi }}{(\zeta \Delta )e^{i\psi }+1}+(\theta -\phi )\frac{
    (\zeta /\Delta )e^{i\psi }}{(\zeta /\Delta )e^{i\psi }+1}\right] \,,
\end{equation}
where following the notations of paper \cite{we} we have introduced the variable $\zeta >0$, the angle $\psi \in \left(
-\pi /2,\pi /2\right) $, and the parameter $\Delta >0$ given by the expression
\begin{equation}\label{ib:3}
    \Delta^2  =\sqrt{\frac{D_{1}}{D_{2}}} \, \frac{\tau _{1}(\chi)}{\tau _{2}(\chi )}
    =\Delta^2_\phi\exp \left\{\theta(1-2\chi) \right\}
    \qquad \text{with} \qquad
    \Delta^2_{\phi}= \sqrt{ \frac{D_{1}}{D_{2}}}\, \frac{\nu _{2}}{\nu _{1}}\exp \left\{\eta - \phi\right\}
\end{equation}
\end{widetext}
such that
\begin{gather}
\label{ib:5}
    \gamma  =\frac{a^{2}}{\sqrt{D_{1}D_{2}}\tau _{1}(\chi )\tau _{2}(\chi )} \,\zeta ^{2}e^{i2\psi},
\\
\intertext{and}
\label{ib:5a}
    p_{1} = \frac{a}{D_{1}\tau _{1}(\chi )}\,\Delta \zeta e^{i\psi }\,,
    \quad
    p_{2} = \frac{a}{D_{2}\tau _{2}(\chi )}\,\frac{1}{\Delta }\zeta e^{i\psi }\,.
\end{gather}
The quantity $g$ stands for the dimensionless diffusion flux of species through the aqueous solution bulk towards the
crystal surface
\begin{multline}\label{ib:6}
    g  = \sqrt{D_{1}D_{2}}\tau _{1}(\chi )\tau _{2}(\chi )G
 \\
     {}  = \sqrt{D_{1}D_{2}}\tau^{2}_g \exp\left\{\phi (1-2\chi )+\theta\right\} G\,.
\end{multline}

To find the boundary of the instability region in the space of system parameters we note that the eigenvalue
equation~(\ref{ib:2}) can be directly reduced to a fourth-order polynomial equation by multiplying it by both the
denominators entering its right-hand side. The coefficient of the highest power term of this polynomial is a constant
value. So the roots of equation~(\ref{ib:2}) cannot go to infinity and, thus, vary continuously as the system
parameters change. The instability boundary separates the regions where the value of $\operatorname{Re}\gamma $ has
different signs and, therefore, meets the equality
\begin{equation*}
    \operatorname{Re}\gamma =0
\end{equation*}
converting, due to \eqref{ib:5}, into the condition $\psi =\pm \pi/4 $. Taking the latter into account and splitting
equation~(\ref{ib:2}) into the real and imaginary parts we immediately get the conclusion that at the instability
boundary the parameter $\zeta$ obeys the following equation
\begin{equation}
\label{ib:7}
    (\theta +\phi)\Psi_1\Big(\zeta\Delta \Big) +
    (\theta -\phi)\Psi_1\Big(\frac{\zeta }{\Delta}\Big)
%    =\frac{1}{\chi(1-\chi )}\,,
    =2\Theta_c
\end{equation}
and the diffusion flux takes the value
\begin{multline}
%    g_{c}=\frac{\zeta }{\sqrt{2}\chi (1-\chi )}
    g_{c}=\sqrt2\Theta_c \zeta_c
\\
    {}\times
    \left[(\theta +\phi)\Delta\Psi_2\Big(\zeta_c\Delta\Big) +
    (\theta -\phi)\frac{1}{\Delta}\Psi_2\Big(\frac{\zeta_c}{\Delta }\Big)\right]^{-1},
\label{ib:8}
\end{multline}
where $\zeta_c$ is the solution of equation~\eqref{ib:7} and the functions
\begin{equation}
    \Psi _1(x) =\frac{\sqrt2x(\sqrt2x+1)}{(\sqrt2x+1)^{2}+1}\,,\
    \Psi _2(x) =\frac{1}{(\sqrt2x+1)^{2}+1}  \label{ib:9}
\end{equation}
as well as the critical value of the nonideality parameter depending on the crystal composition $\chi$
\begin{equation}
\label{ibn:14b}
    \Theta_c(\chi) =\frac{1}{2\chi(1-\chi)}\,.
\end{equation}
have been introduced. In other words, at the instability boundary the general eigenvalue equation~\eqref{ib:2} is reduced to \eqref{ib:7} and if its solution $\zeta_c$ exists then formula~\eqref{ib:8} specifies the critical value of the species diffusion flux $g_c$. Only one additional condition should be imposed; it is the requirement that the obtained value of $g_c$ be positive.

Below we will confine our consideration to the case $\chi=0.5$ only for which $\Theta_c:=
\theta_{c0} = 2$. It due to, first, exactly this value of the solid composition $\chi$ determines actually the
boundaries, external and internal ones, of the instability regions to be analyzed. Second, as follows directly from
expressions~\eqref{ib:7} and \eqref{ib:8}, by transformations
\begin{align}\label{new:7}
    \theta_\text{new} &= \theta_\text{old}\cdot\frac{\Theta_c}{\theta_{c0}} &
    \phi_\text{new} &= \phi_\text{old}\cdot \frac{\Theta_c}{\theta_{c0}} &
\end{align}
the case of $\chi\ne 0.5$ is reduced immediately to the given one. Naturally the dependence of the system characteristics on the solid composition $\chi$ endows the growth instabilities with nontrivial properties. In particular, in some sense ``optimal'' conditions of the instability onset can match the solid composition deviating substantially from $\chi = 0.5$, which in turn is able to cause a system instability with respect to spatially nonuniform perturbations. This question, however, is beyond the scope of the present paper. Third, in the mathematical expressions to be obtained below the quantity $\theta_{c0}$ will be kept on instead of being replaced by its numerical value, so using transformations~\eqref{new:7} the general expressions can be reconstructed immediately.

The solution of the system \eqref{ib:7} and \eqref{ib:8} implicitly determines the critical value
$g_{c}(\theta,\phi,\Delta)$ of the species diffusion flux. Thereby it describes the boundary of the instability region
in the complete space of the system parameters $\{g,\theta,\phi,\Delta\}$. Projecting this region onto various planes
makes it possible to regard the instability boundary as some curve (or surface) dividing a given plane into two
domains, where the instability can arise in principle for a given values of the corresponding parameters or cannot do
it at all. Below in this Section we will consider in detail this projection onto the plane $\{\theta,\phi\}$ for a
fixed value of the parameter $\Delta$ with the main attention paid to the limit  $\Delta\gg1$.

\subsection{Two mechanisms of the instability}

\begin{figure}
\begin{center}
\rotatebox{90}{\hspace{1.25cm} LHS of equation \eqref{ib:7}}
\includegraphics[width=56mm]{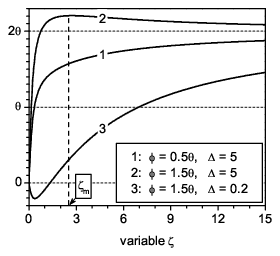}
\end{center}
\caption{Left-hand side (LHS) of equation \eqref{ib:7} as a function of the variable $\zeta$. Curve 1 depicts this
dependence when the system asymmetry cannot affect the instability onset crucially ($\phi<\theta$), curve 2 exhibits
the case where the asymmetry effect is pronounced ($\phi > \theta$). Curve 3 demonstrates the fact that the system asymmetry depresses the instability onset when $\phi>\theta$ and $\Delta <1$.}
\label{Fig.4}
\end{figure}

Possible roots $\{\zeta_c\}$ of equation~\eqref{ib:7} specify the eigenvalues determining the critical value of the diffusion flux $g_c$ via expression~\eqref{ib:8}. The instability boundary is the locus where the potentials $\phi$ and $\theta$ take such values that the left-hand side of equation~\eqref{ib:7} gets its maximum at these roots. The solid nonideality and the system asymmetry are responsible for the terms in this equation exhibiting different behavior. The term proportional to $\theta$ is the increasing function of $\zeta$, whereas one proportional to $\phi$ comprises increasing and decreasing branches. It is the mathematical reflection of different instability mechanisms caused by  the solid nonideality and the system asymmetry.

The asymmetry effect becomes crucial when the left-hand side of equation~\eqref{ib:7} changes its behavior as a function of $\zeta$. It converts from a function monotonically increasing from 0 to $2\theta$ when $\zeta$ runs from 0 to $\infty$ (curve 1 in Fig.~\ref{Fig.4}) to one possessing a maximum $\Psi_m$ attained at a certain internal point $0<\zeta_m <\infty$ (curve 2 in Fig.~\ref{Fig.4}). For $0<\zeta<\zeta_m$ it grows from 0 to $\Psi_m > 2\theta$ and then drops down to $2\theta$ on the interval $\zeta_m < \zeta<\infty$. The asymptotics of the left-hand side of equation~\eqref{ib:7} as $\zeta\to\infty$ demonstrates us directly that it is the case when
\begin{equation}\label{ib:11}
    \left(
    \frac{\phi -\theta }{\phi +\theta }
    \right)
    \Delta^2 > 1\,.
\end{equation}
In fact, if inequality~\eqref{ib:11} holds the asymptotics of the left-hand side of equation~(\ref{ib:7}) is a decreasing function of $\zeta$ and, thus, the point $0<\zeta _{m}<\infty $ does exist. Exactly in this case the instability can arise even the nonideality potential is less then its threshold, $\theta < \theta_{c0}$ provided the maximum $\Psi_m > 2\theta_{c0}$ due to the effect of the system asymmetry. For the latter to be the case the inequality $\Delta >1$ is necessary as follows from condition~\eqref{ib:11}. For $\Delta<1$ the system asymmetry depresses the instability onset as it is illustrated in Fig.~\ref{Fig.4} by curve~3.

\begin{figure}
\begin{center}
\includegraphics[width=70mm]{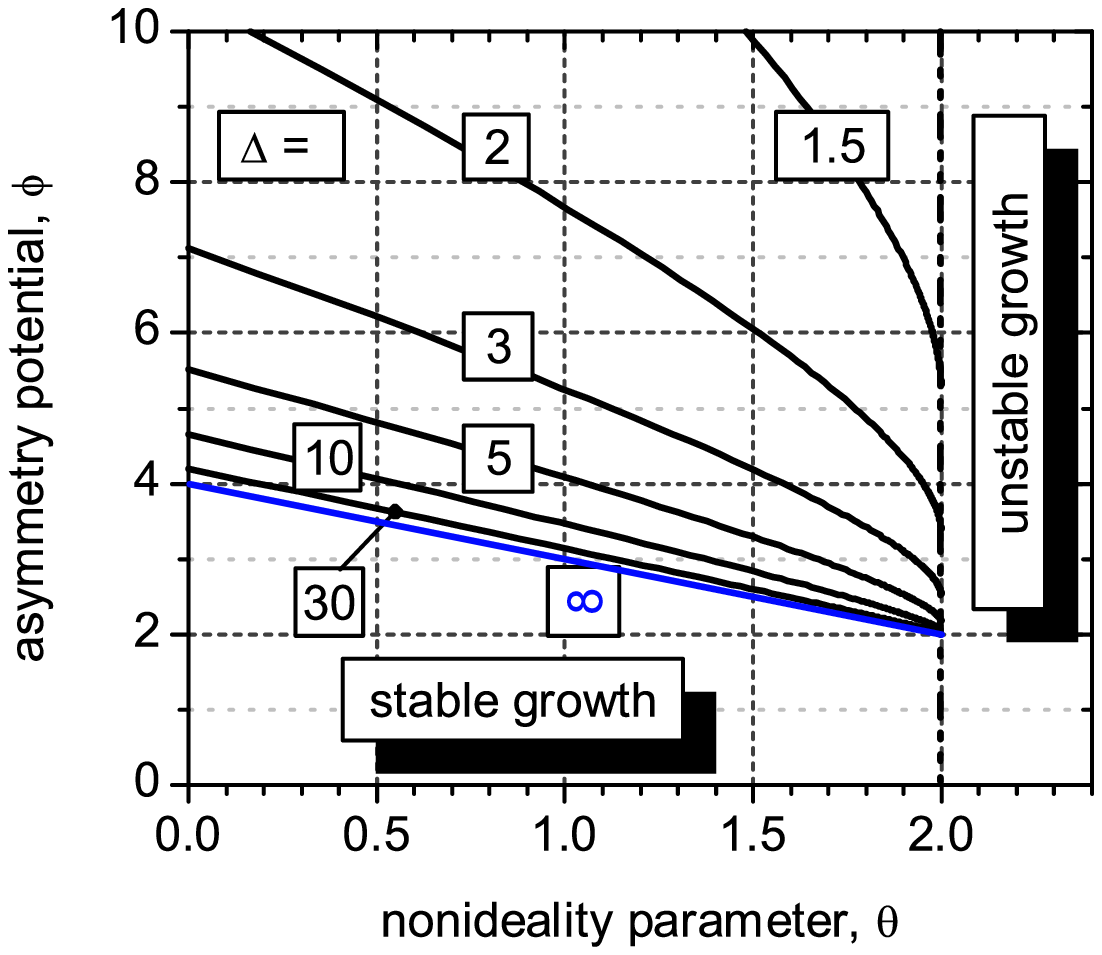}
\end{center}
\caption{The instability boundary $\mathcal{B}_\phi^+$ for several values of the parameter $\Delta$ including the limit value $\Delta=\infty$. The used criterion of instability is $\text{Re\,}\gamma>0$ for $g\to\infty$.}
\label{Fig.13}
\end{figure}

Before passing to a detailed analysis of the instability domain we present a fairly simple way to construct the instability boundary of the plane $\{\theta,\phi\}$ for a fixed value of $\Delta$. It applies to the fact that the given system admits two scenarios of the instability onset. One caused by the solid nonideality matches the eigenvalues $\zeta e^{i\psi}\to\infty$ with the diffusion flux $g\to\infty$. In this case the solution of the general eigenvalue equation~\eqref{ib:2} can be written as
\begin{align}\label{verynew:1}
    \zeta^2 e^{i2\psi} & = g\left(\frac{\theta}{\theta_{c0}}-1\right)
    &&
    \text{for}
    &&
    g\to\infty\,,
\end{align}
so the instability arises when the nonideality parameter exceeds its critical value, $\theta >\theta_{c0}$, because $\gamma\propto\zeta^2e^{i2\psi}$. The other is characterized by the bounded variations of the eigenvalues $\zeta e^{i\psi}$ as the diffusion flux goes to infinity. Under this condition we can analyze directly the eigenvalue problem in the limit $g\to\infty$ setting the left-hand side of equation~\eqref{ib:2} equal to zero and, thus, reducing it actually to a quadratic equation. Omitting simple arithmetical manipulations the result is
\begin{gather}
\label{ib:90}
    \zeta e^{i\psi} = \frac1{4(\theta_{c0} -\theta)}\left[
    \kappa \pm \sqrt{\kappa^2 - 16\theta_{c0}(\theta_{c0}-\theta)}
    \right]\,,
\\
\intertext{where}
\label{ib:91}
    \kappa := \left(\Delta + \frac1\Delta\right)\theta + \left(\Delta - \frac1\Delta\right)\phi
    - 2\left(\Delta + \frac1\Delta\right)\theta_{c0}\,.
\end{gather}
For $\theta >\theta_{c0}$ one of these roots corresponds to unstable perturbations, nevertheless, the perturbations matching the eigenvalues given by expression~\eqref{verynew:1} are dominant due to large values of their increments. However, when the solid nonideality is not to high, i.e. $\theta<\theta_{c0}$, the latter perturbations turn out to be stable and the growth instability is caused by the system asymmetry. Indeed, the instability boundary with $\psi = \pm\pi/4$ meets the condition
\begin{align}\label{ib:92}
    \kappa & >0 & &\text{and}&
    \kappa^2 &= 8\theta_{c0}(\theta_{c0}-\theta)\,.
\end{align}
By virtue of \eqref{ib:92} such instability can arise when the parameters $\phi$ and $\Delta >1$ reflecting the system asymmetry meet the inequality
\begin{equation}\label{verynew:2}
    \phi > \phi^+_{c} = \frac{\Delta^2+1}{\Delta^2-1}\left(2\theta_{c0} - \theta\right)
    + \frac{2\sqrt2\Delta}{\Delta^2-1}\sqrt{\theta_{c0}(\theta_{c0}-\theta)}\,.
\end{equation}
When $\Delta < 1$ the system asymmetry suppresses the instability as noted above applying to Fig.~\ref{Fig.4}. The curve $\mathcal{B}_\phi^+$ on the plane $\{\theta,\phi\}$ specified by the dependence $\phi_c^+(\theta)$ is presented in Fig.~\ref{Fig.13} for several values of the parameter $\Delta$. Roughly speaking $\mathcal{B}_\phi^+$ is the boundary of the instability domain for $\theta < \theta_{c0}$.

It should be underlined that the present analysis was based on the assumption that the instability has to arise for large values of the species diffusion flux if it can develop in principle for given values of the other system parameters. It is true when the growth instability is caused by the solid nonideality. However, for the instability induced by the system asymmetry the situation is more intricate. Rigorously speaking, in the latter case at the real instability boundary $\mathcal{B}_\phi$ the species diffusion flux $g$ takes a certain finite value $g_{c}<\infty$ and in a narrow boundary layer inside the instability region the diffusion flux must belong to a finite interval, $g_c<g<g^+_c<\infty$. Nevertheless, as will be seen below, this feature is valuable only for $\Delta\gtrsim1$. So as stems from expression~\eqref{verynew:2} for $\Delta\gg1$ the boundary of the growth instability caused by the system asymmetry is approximated by the line
\begin{equation}\label{verynew:3}
    \phi =  2\theta_{c0} - \theta
\end{equation}
at the leading order in $1/\Delta$.

\section{Structure of the instability domain}

\begin{figure}
\begin{center}
\includegraphics[width = 60mm]{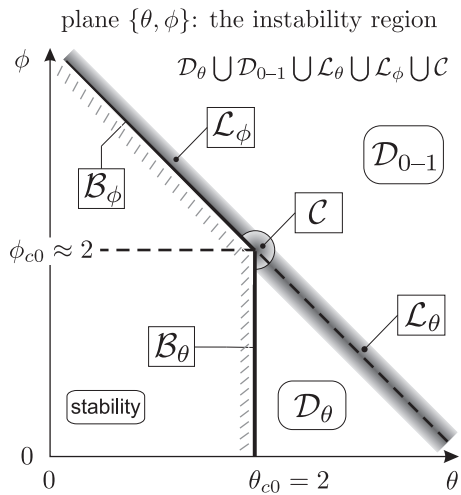}
\end{center}
\caption{The structure of the instability region as a whole on the plane $\{\theta,\phi\}$ for a fixed value of the
parameter $\Delta\gg1$. It comprises five regions distinguishable in properties: two volumetric domains
$\mathcal{D}_\theta$ and $\mathcal{D}_\text{0--1}$, one intermediate layer $\mathcal{L}_\theta$ between them and one
boundary layer $\mathcal{L}_\phi$ whose thickness $W_c\gtrsim1/\Delta\ll1$, and, finally, a double criticality
neighborhood  $\mathcal{C}$ of the point $\{\theta_{c0},\phi_{c0}\approx\theta_{c0} \}$.}
\label{Fig.7a}
\end{figure}

Based on the analysis of the eigenvalue equation~\eqref{ib:7} for $\Delta\gg1$ we can single out five characteristic
regions of the system instability on the plane $\{\theta,\phi\}$ shown in Fig.~\ref{Fig.7a}. Let us consider them
individually assuming $\Delta\gg1$ to hold.

\subsection{Instability domain $\mathcal{D}_\theta$}

The volumetric domain $\mathcal{D}_\theta$ matches actually the growth instability studied in our previous paper \cite{we}. It is bounded by the vertical line $\mathcal{B}_\theta = \{\theta,\phi : \theta = \theta_{c0}=2\}$, by the layer $\mathcal{L}_\theta$, and the $\theta$-axis. The layer $\mathcal{L}_\theta$ is a certain neighborhood of the line $\phi = 2\theta_{c0} - \theta$ whose thickness is about $W_c \gtrsim 1/\Delta$. In domain $\mathcal{D}_\theta$ condition~\eqref{ib:11} is strongly violated, i.e.
\begin{equation}\label{ib:11D1}
    \left(
    \frac{\theta-\phi}{\phi +\theta }
    \right)
    \Delta^2 \gg 1\,.
\end{equation}
So the left-hand side of equation~\eqref{ib:7} is monotonically increasing function of $\zeta$. Besides, for any point of the domain $\mathcal{D}_\theta$ the distance between it and the line $\phi = 2\theta_{c0} - \theta$, i.e. actually between it and the layer $\mathcal{L}_\theta$, can be regarded as a large value in comparison with the quantity $1/\Delta$. The latter statement, as can be shown directly, causes the solution of equation~\eqref{ib:7} to meet the inequality $\zeta_c\gg1$. Thereby the former term on the left-hand side of equation~\eqref{ib:7} can be taken in the limit $\zeta\Delta\to\infty$. Also the corresponding term in expression~\eqref{ib:8} is ignorable. Under these conditions equation~\eqref{ib:7} is reduced to a quadratic equation with respect to $\zeta$, yielding us immediately its solution in the form
\begin{equation}
\label{ibn:12a}
    \zeta_c =\frac{\Delta}{\sqrt2}\, F(r_\theta) \,.
\end{equation}
Here, by definition, the function $F(r_\theta)$ is determined by the expression
\begin{gather}
\label{ibn:13}
    F(x)  = \frac1{2(1-x)}\left[
    2x-1 +\sqrt{1+4x(1-x)}
    \right]\,,
\\
\intertext{its argument is}
\label{ibn:14a}
    r_\theta  = \frac{2\theta_{c0} - \theta-\phi }{\theta-\phi}
    \equiv 1-\frac{2}{(\theta-\phi)}\,\left[\theta -\theta_{c0}\right]\,.
\end{gather}
The inequality  $\theta_{c0} <\theta$ is assumed to hold, thus, $0< r_\theta < 1$. Then the critical value $g_c$ of the
dimensionless diffusion flux is
\begin{equation}
\label{ibn:15}
    g_{c\{\mathcal{D}_\theta\}}  = \frac{ \Delta^2 \theta_{c0}}{(\theta-\phi)} \,\frac{F^2(r_\theta)\left[F(r_\theta)+1\right]}{r_\theta}
\end{equation}
by virtue of \eqref{ib:8}.

Near the threshold of the nonideality coefficient, i.e. in the vicinity of the boundary $\mathcal{B}_\theta$, where
\begin{equation}\label{new:0}
     0< \theta - \theta_{c0} \ll \theta - \phi\,,
\end{equation}
one has $1 -r_\theta \ll 1$. In this case function~\eqref{ibn:13} is approximated as $F(r_\theta)\approx
1/(1-r_\theta)$ and expression~\eqref{ibn:15} is reduced to
\begin{equation}\label{ibn:15h}
    g_{c\{\mathcal{D}_\theta|\mathcal{B}_\theta\}}  \approx \frac{\Delta^2\theta_{c0}\,(\theta-\phi)^2}{8(\theta-\theta_{c0})^3}\,.
\end{equation}
Whence it follows, in particular, that the critical value of diffusion flux diverges as $(\theta-\theta_{c0})^{-3}$ for
$\theta\to\theta_{c0}+0$, being in agreement with the results of paper~\cite{we}.

When the analyzed point $\{\theta,\phi\}$ is located in a close proximity to the layer $\mathcal{L}_\theta$, i.e. for
\begin{equation}\label{new:1}
       \frac1{\Delta} \ll 2\theta_{c0} -\theta -\phi \ll \theta - \phi
\end{equation}
one has $r_\theta \ll 1$ and $F(r_\theta)\approx 2r_\theta$. In this case formula~\eqref{ibn:15} is
simplified as
\begin{equation}\label{ibn:15hh}
    g_{c\{\mathcal{D}_\theta|\mathcal{L}_\theta\}}  \approx \frac{4\Delta^2\theta_{c0}\,(2\theta_{c0}-\theta-\phi)}{(\theta-\phi)^2}\,.
\end{equation}
We remind that in expression~\eqref{ibn:15hh} the difference $(2\Theta_{c}-\theta-\phi)$ cannot become too small
according to inequality~\eqref{new:1}. The behavior of the critical diffusion flux for points coming close to the line
$\phi= 2\theta_{c0}-\theta$ is considered below.

\subsection{Instability domain $\mathcal{D}_\text{0--1}$}

The other volumetric domain $\mathcal{D}_\text{0--1}$ of system instability is formally the half-plane bounded from
below by the layer composition $\mathcal{L}_\theta\bigcup\mathcal{L}_\phi$, i.e. by a neighborhood of the line $\phi =
2\theta_{c0}-\theta$ with thickness $W_c\gtrsim 1/\Delta$ (Fig.~\ref{Fig.7a}). It comprises all the points
$\{\theta,\eta\}$ such that
\begin{equation}\label{new:2}
    \theta + \phi - 2\theta_{c0}\gg \frac1{\Delta}\,.
\end{equation}
In this region the solution $\zeta_c$ of equation~\eqref{ib:7} turns out to be much less than unity, $\zeta_c\ll 1$. As
can be verified directly the latter inequality enables us to ignore both the second terms on the left-hand side of
equation~\eqref{ib:7} and inside the square brackets in equation~\eqref{ib:8}. The appearance of these terms is due to
the channel 0--2 of the precipitation reactions~\eqref{Intr:5}. Therefore in the domain $\mathcal{D}_\text{0--1}$ the
contribution of the channel~0--2 is of minor importance and the growth instability is caused by the channel 0--1
individually.

Using this simplification the eigenvalue equation~\eqref{ib:7} again can be reduced to a quadratic equation with the
solution
\begin{align}
\label{ibn:20}
    \zeta_c & = \frac1{\sqrt2\,\Delta}\, F(r_{01}) \,,
\\
\intertext{where, by definition, the argument $r_{01}$ is the value}
\label{ibn:21}
    r_{01} & = \frac{2\theta_{c0}}{\theta+\phi}
\end{align}
and meets the inequality $1-r_{01} \gg 1/\Delta$.

The corresponding expression for the critical value of the species diffusion flux is
\begin{equation}\label{ibn:22}
    g_{c\{\mathcal{D}_\text{0--1}\}}  = \frac1{\Delta^2}\,F^2(r_{01})\left[F(r_{01})+1\right]\,.
\end{equation}
In particular, near the domain boundary  $\mathcal{L}_\theta\bigcup\mathcal{L}_\phi$, i.e. for
\begin{equation}\label{new:4}
   \frac1\Delta \lesssim  \theta +\phi - 2\theta_{c0} \ll 1\,,
\end{equation}
where $1-r_{01}\ll 1$ and the function $F(r_{01})\approx 1/(1-r_{01})$ expression~\eqref{ibn:22} converts into
\begin{equation}
\label{ibn:23}
    g_{c\{\mathcal{D}_\text{0--1}|\mathcal{L}_\theta\bigcup\mathcal{L}_\phi\}}  \approx \frac{8\theta_{c0}^3}{\Delta^2 (\theta +\phi - 2\theta_{c0} )^{3}}\,.
\end{equation}
It should be pointed out that expression~\eqref{ibn:23} does not describe a real singularity in the diffusion flux
threshold. In fact, the difference
\begin{equation}\label{new:3}
    \sigma := \theta+\phi - 2\theta_{c0}
\end{equation}
is bounded from below in the domain $\mathcal{D}_\text{0--1}$, namely, $\sigma \gg 1/\Delta$ and the limit $\sigma\to
+0$ cannot be implemented in it.

By virtue of expression~\eqref{ibn:22} in the domain $\mathcal{D}_\text{0--1}$ the diffusion flux threshold $g_c$
practically does not depend on the particular value of the difference $(\theta-\phi)$ because of the minor effect of
channel~0--2. The situation changes dramatically when the analyzed point $\{\theta,\phi\}$ enters the boundary of this
domain, being the subject of the following subsections.

\subsection{Intermediate layer $\mathcal{L}_\theta$}

\begin{figure}
\begin{center}
\includegraphics[width=60mm]{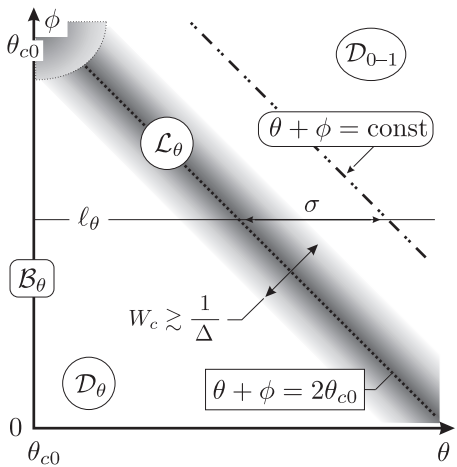}
\end{center}
\caption{The intermediate layer $\mathcal{L}_\theta$ separating the instability domains $\mathcal{D}_\theta$ and
$\mathcal{D}_\text{0--1}$. Zoomed-in view.}
\label{Fig.7b}
\end{figure}

The instability domains $\mathcal{D}_\theta$ and $\mathcal{D}_\text{0--1}$ are joined to each other via the layer
$\mathcal{L}_\theta$ whose points are located near the line $\theta+\phi = 2\theta_{c0}$ and meet the inequality
$\theta>\phi$. So the left-hand side of the eigenvalue equation~\eqref{ib:7} is a monotonously increasing function of
$\zeta$ and the solution of this equation $\zeta_c$ decreases as the potential $\theta$ increases for a fixed value of
$\phi$. The results obtained in the two previous subsections show us that the quantity $\zeta_c$, first, drops from
very large values up to $\zeta_c\gtrsim 1$ as the analyzed point $\{\theta,\phi\}$ goes from the instability boundary
$\mathcal{B}_\theta$ to the layer $\mathcal{L}_\theta$. Then, just after the point crossing the layer
$\mathcal{L}_\theta$ the quantity $\zeta_c$ takes values about $\zeta_c\lesssim 1$ and drops down to zero as the
analyzed point goes away from it. In fact, on one hand, by virtue of \eqref{ibn:12a} when the analyzed point
$\{\theta,\phi\}$ tend to the layer $\mathcal{L}_\theta$ on the side of the domain $\mathcal{D}_\theta$ and
inequality~\eqref{new:1} holds we have
\begin{equation*}
        \zeta_c = \frac{\sqrt2 \Delta (2\theta_{c0} -\theta -\phi)}{\theta - \phi}\,.
\end{equation*}
On the other hand, for the point $\{\theta,\phi\}$ located near the layer $\mathcal{L}_\theta$ on the side of the
domain $\mathcal{D}_\text{0--1}$  where the inequality~\eqref{new:4} hold the solution $\zeta_c$ of the eigenvalue
equation~\eqref{ib:7} is approximated as
\begin{equation*}
  \zeta_c = \frac{\sqrt2\theta_{c0}}{\Delta(\theta+\phi-2\theta_{c0})}
\end{equation*}
by virtue of \eqref{ibn:20}. The ``boundaries'' of the layer $\mathcal{L}_\theta$ meet the estimate $\Delta|\theta
+\phi-2\theta_{c0}|\gtrsim1$, which justifies the statement mentioned above. So inside the layer $\mathcal{L}_\theta$
the quantity $\zeta_c$ has to change in the interval $1 \lesssim \zeta_c \lesssim 1$.

The expression obtained below for the critical value of the diffusion flux $g_c$ is valid, however, for a wider region
than the layer $\mathcal{L}_\theta$ itself due to the adopted assumption $\Delta\gg1$. Namely, in this subsection we
consider the region for which
\begin{gather}
\label{new:5a}
         |\theta+\phi - 2\theta_{c0}| \ll 1\,,
\\
\intertext{thereby the two inequalities}
\label{new:5b}
        \zeta_c\Delta \gg 1\quad\text{and}\quad \frac{\zeta_c}{\Delta} \ll 1\,.
\end{gather}
hold simultaneously. This region comprises the layer $\mathcal{L}_\theta$ as well as the neighboring parts of the
domains $\mathcal{D}_\theta$ and $\mathcal{D}_\text{0--1}$. So the expression for the diffusion flux threshold valid in
it really specifies the crossover between the domains $\mathcal{D}_\theta$ and $\mathcal{D}_\text{0--1}$.

Under condition~\eqref{new:5b} the former term on the left-hand side of equation~\eqref{ib:7} can be approximated by
the asymptotics of the function $\Psi_1(x)$ for $x\to\infty$, whereas the latter one matches the limit $x\to 0$.
Therefore in this case equation~~\eqref{ib:7} can be rewritten as
\begin{equation}\label{ibn:25}
    (\theta + \phi)\frac1{\zeta} - (\theta-\phi)\zeta = \sqrt2\Delta \sigma \,,
\end{equation}
where $\sigma$ is given by expression~\eqref{new:3}.  The solution of \eqref{ibn:25} is of the form
\begin{equation}\label{ibn:26}
    \zeta_c  = \frac1{\sqrt2 (\theta - \phi)}
    \left[\sqrt{\Delta^2\sigma^2 + 2(\theta^2 - \phi^2)} -\Delta\sigma \right]\,.
\end{equation}
Then the substitution of \eqref{ibn:26} into \eqref{ib:8} yields
\begin{equation}\label{ibn:27}
    g_{c\{\mathcal{L}_\theta\}}  = \frac{2\theta_{c0}\Delta\zeta_c^2}{\sqrt{\Delta^2\sigma^2 + 2(\theta^2 - \phi^2)}}
\end{equation}
where the function $\Psi_2(x)$ has been also approximated using the appropriate asymptotics. As it must be
expression~\eqref{ibn:27} converts into expressions~\eqref{ibn:15hh} and \eqref{ibn:23} for $\mp\Delta\sigma\gg1$,
respectively.

\begin{figure}
\begin{center}
\includegraphics[width=\columnwidth]{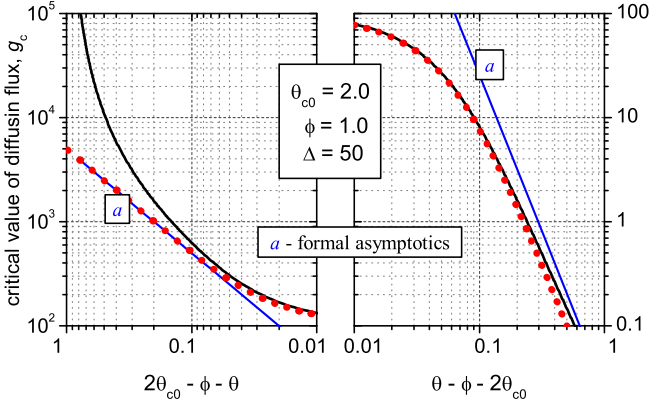}
\end{center}
\caption{The critical value of the dimensionless diffusion flux $g_c$ vs the nonideality parameter $\theta$ near the
intermediate layer $\mathcal{L}_\theta$. The plot is based on the general equations~\eqref{ib:7} and \eqref{ib:8} using the parameters shown in inset. The straight lines visualize the formal asymptotics~\eqref{ibn:15hh} and \eqref{ibn:23} whereas the dotted line corresponds to \eqref{ibn:27}.}
\label{Fig.6}
\end{figure}

Figure~\ref{Fig.6} illustrates the obtained crossover of the diffusion flux threshold. In this Figure the critical value $g_c$ of the dimensionless diffusion flux is shown vs actually the nonideality parameter $\theta$ for fixed parameters $\phi = 1$ and $\Delta = 50$. In other words, it visualizes $g_c(\theta)$ for the analyzed point $\{\theta,\phi\}$ moving along the line $\ell_\theta$ shown in Fig.~\ref{Fig.7b}.

\subsection{Boundary layer $\mathcal{L}_\phi$}

\begin{figure}
\begin{center}
\includegraphics[width=60mm]{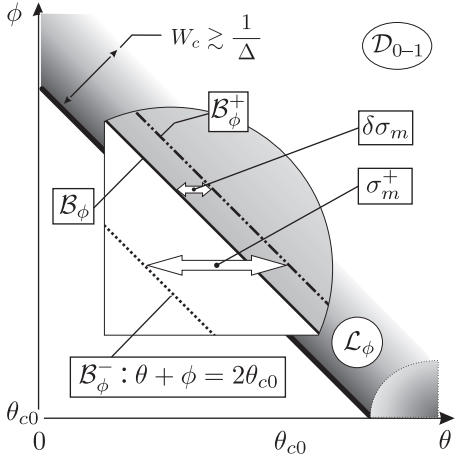}
\end{center}
\caption{The boundary layer $\mathcal{L}_\theta$ separating the instability domain $\mathcal{D}_\text{0--1}$ and the
region of the stable crystal growth. Zoomed-in view with an additional magnifying lens showing the finite structure of
the instability boundary $\mathcal{B}_\phi$.}
\label{Fig.7c}
\end{figure}

When the potential $\theta$ is less than the threshold $\theta_{c0}$ the solid nonideality cannot individually cause
the system instability. In this case only the cumulative effect of the system nonideality and asymmetry gives rise to
the growth instability or even the system asymmetry itself does when the potential $\phi$ is high enough. So for
$\theta <\theta_{c0}$  the instability domain $\mathcal{D}_\text{0--1}$ borders with the region of the stable crystal
growth via the boundary layer $\mathcal{L}_\phi$ (Fig.~\ref{Fig.7c}). Let us consider its properties in detail. As for
the layer $\mathcal{L}_\theta$ analyzed in the previous subsection the given layer matches the root $\zeta_c$ of
equation~\eqref{ib:7} of order unity, $\zeta_c\sim 1$. However in this case by virtue of condition~\eqref{ib:11} the
potential $\phi$ should exceed the nonideality parameter, $\phi>\theta$, for the growth instability to arise. As a
result the two terms entering the left-hand side of equation~\eqref{ib:7} have opposite signs and the functions
$\Psi_1(x)$, $\Psi_2(x)$ should be approximated to the next order in the corresponding small parameters in comparison
with the case of the layer $\mathcal{L}_\theta$. Namely, using inequalities~\eqref{new:5b} expressions~\eqref{ib:7} and
\eqref{ib:8} are reduced to the equation
\begin{equation}
\label{ib:70}
    \frac{(\phi+\theta)}{\zeta} + (\phi - \theta)\zeta
     = \sqrt2\Delta\sigma
     + O\left(\frac1{\Delta^{2}}\right)
\end{equation}
and the expression for the diffusion flux threshold
\begin{equation}\label{ib:71}
    g_{c}  =  \frac{2\sqrt2\theta_{c0}\Delta\zeta^2}{\Phi(\zeta)}\,,
\end{equation}
where the function $\Phi(\zeta)$ is introduced by the formula
\begin{multline}
\label{ib:72}
    \Phi(\zeta):=
    \bigg[\frac{(\phi+\theta)}{\zeta} -(\phi-\theta)\zeta\bigg]
\\
     {} - \frac{\sqrt2}{\Delta}
    \bigg[\frac{(\phi+\theta)}{\zeta^2} -(\phi-\theta)\zeta^2\bigg]+
    O\bigg(\frac{1}{\Delta^2} \bigg)\,.
\end{multline}
It should be pointed out that expression~\eqref{ib:70} does not contain a term of order $\Delta^{-1}$ and the other
term of order $\Delta^{-2}$ is not written explicitly because its effect is reduced only to a small constant
contribution to the value of $\sigma$.

\begin{figure}
\begin{center}
\includegraphics[width=\columnwidth]{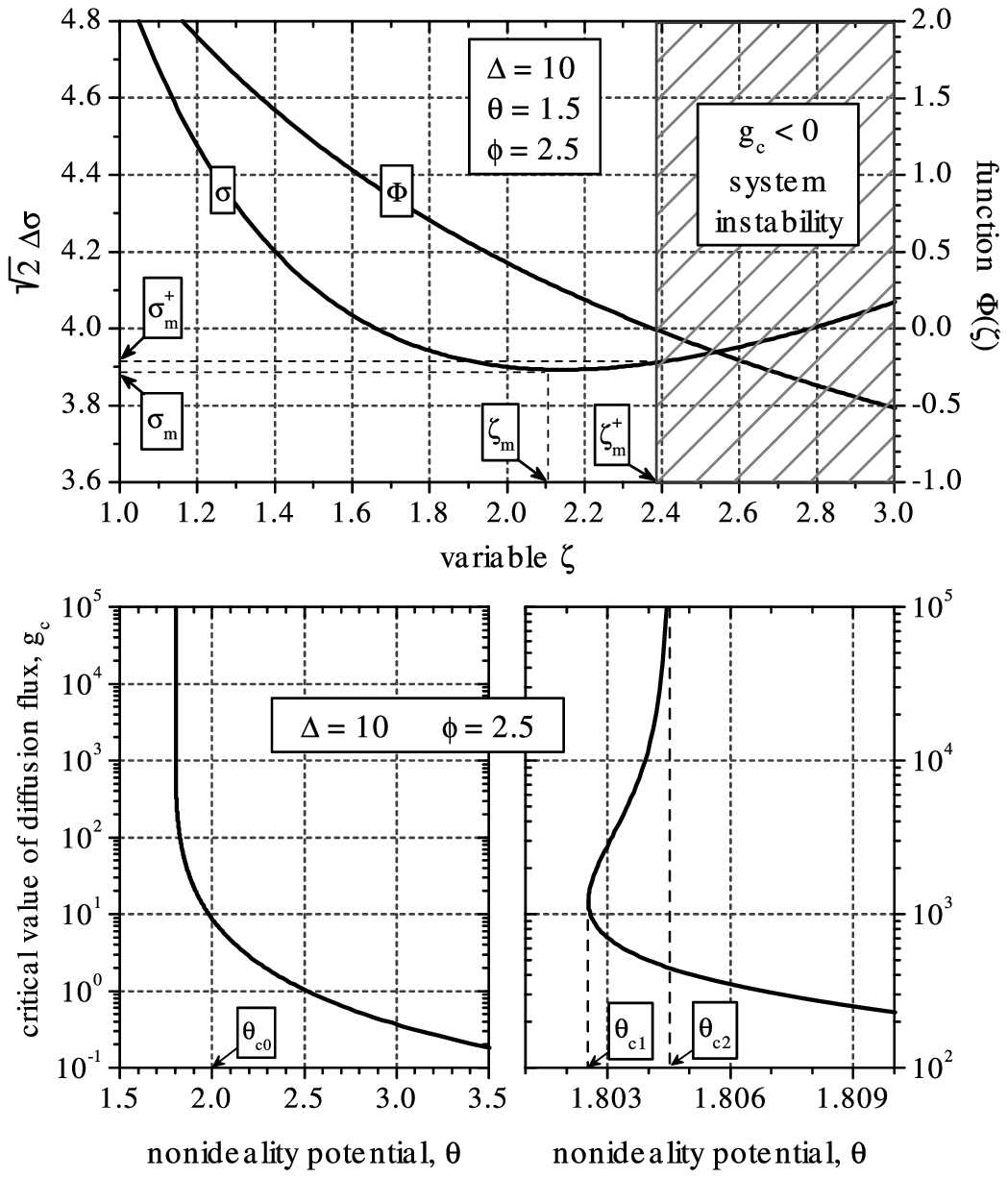}
\end{center}
\caption{Illustration of the mechanism responsible for the complex behavior of the critical diffusion flux in the boundary layer $\mathcal{L}_\phi$ (upper fragment) and the resulting $g_c(\theta)$-dependence (lower fragments). The right-hand fragment depicts this dependence in zoom, making it evident that in the region $(\theta_{c1},\theta_{c2})$  the growth rate should belong to a bounded interval for the instability to arise.
In plotting the potential difference $\sigma$ as a formal function of the variable $\zeta$ determined by equation~\eqref{ib:70} and the function $\Phi(\zeta)$ (see equation~\eqref{ib:71}) vs the variable $\zeta$ the values $\theta = 1.5$, $\phi = 2.5$, and $\Delta = 10$ were used as well as the original functions $\Psi_1(x)$ and $\Psi_2(x)$ (expression~\eqref{ib:9}) rather then their approximations were applied to take into account not only the leading terms but also all the other small contributions.}
\label{Fig.5}
\end{figure}

To explain the resulting dependence of the diffusion flux threshold $g_c$ on the potential difference $\sigma$ let us
refer to Fig.~\ref{Fig.5}. It illustrates the value of $\sigma$ treated as a formal function of $\zeta$ that is
determined by equation~\eqref{ib:70} and has a minimum $\sigma_m$ meeting the estimate
\begin{align}
\label{ib:73}
    \Delta\sigma_m & = \sqrt{2(\phi^2-\theta^2)} + O\left(\frac1{\Delta^2}\right)
\\
\intertext{and attained at}
\label{ib:74}
    \zeta_m & = \sqrt{\frac{\phi+\theta}{\phi - \theta}} +  O\left(\frac1{\Delta^2}\right)\,.
\end{align}
It matches the critical value of the species diffusion flux
\begin{equation}
\label{ib:75}
    g_{m}  = \frac{\theta_{c0}\Delta^2}{\theta}\left(\frac{\phi+\theta}{\phi-\theta}\right)
\end{equation}
written in the leading order of $1/\Delta$.

If the potential difference $\sigma = \phi + \theta -2\theta_{c0}$ is less than $\sigma_m$ there is no solution of
equation~\eqref{ib:70}  (i.e. of equation~\eqref{ib:7}) and the crystal growth is stable. When $\sigma >
\sigma_m$ equation~\eqref{ib:70} admits two solutions written again in the leading order of $1/\Delta$  as
\begin{align}
\label{ib:76a}
    \zeta_c^- &= \sqrt2(\phi+\theta)\left[
        \Delta\sigma + \sqrt{\Delta^2\sigma^2 - 2(\phi^2 - \theta^2)}
    \right]^{-1}
\\
\intertext{and}
\label{ib:76b}
    \zeta_c^+ &= \frac{1}{\sqrt2(\phi-\theta)}\left[
        \Delta\sigma + \sqrt{\Delta^2\sigma^2 - 2(\phi^2 - \theta^2)}
    \right]\,.
\end{align}
Solution~\eqref{ib:76a} matches the decreasing branch of the dependence $\sigma(\zeta)$ (Fig.~\ref{Fig.5}) and
describes the lower boundary of the diffusion flux threshold $g_c(\sigma)$ obeying the estimate
\begin{equation}\label{ib:77}
    \frac{g_{m}}{g_{c}(\sigma)}
     \approx  \left(1+\frac{\Delta^2}{2\theta}\sqrt{\sigma^2-\sigma_m^2}\right)
    \left(\frac{\sigma + \sqrt{\sigma^2 - \sigma^2_m}}{\sigma_m}\right)^2.
\end{equation}
This branch actually specifies the minimal value $\theta_{c1}(\phi,\Delta)$ of the nonideality
parameter $\theta$ necessary for the growth instability to arise for given values of the parameters $\phi$ and
$\Delta$, namely, by virtue of \eqref{ib:73}
\begin{equation}\label{ib:78}
    \theta_{c1} \approx 2\theta_{c0} - \phi +\frac{2\sqrt2}{\Delta} \sqrt{\phi - \theta_{c0}}\,.
\end{equation}
As it must be expression~\eqref{ib:77} converts into expression~\eqref{ibn:23} for $\Delta\sigma\gg 1$ describing the
behavior of the diffusion flux threshold in the instability domain $\mathcal{D}_\text{0--1}$ near the boundary layer
$\mathcal{L}_\phi$.

Solution~\eqref{ib:76b} describes the upper boundary of the instability region $g_c^+(\sigma)$ which, however, exists only within a rather narrow interval of the potential difference $\sigma$, i.e. when $\sigma_m < \sigma < \sigma_m^+$ (Fig.~\ref{Fig.5}). The parameter $\sigma_m^+$ and the corresponding value $\zeta_m^+$ match the point where the function $\Phi(\zeta)$ changes its sign passing through zero. Exactly at this point the upper branch $g_c^+(\sigma)$ of the diffusion flux threshold goes to infinity and for $\sigma > \sigma_m^+$, i.e. for $\zeta > \zeta_m^+$ it does not exist. In this case the values of the species diffusion flux corresponding to the instability onset are bounded only from below by the threshold $g_{c}(\sigma)$. According to expression~\eqref{ib:70}--\eqref{ib:71} the difference between $\zeta_m$ and $\zeta_m^+$ is a value of the first order in the parameter $1/\Delta$, namely,
\begin{equation}\label{ib:79}
    \zeta_m^+ -\zeta_m = \frac{\sqrt2\,\theta}{\Delta(\phi-\theta)}
\end{equation}
and as a result the corresponding difference of the parameters is
\begin{equation}\label{ib:80}
    \delta\sigma_m := \sigma_m^+ -\sigma_m = \frac{\sqrt2\,\theta^2}{\Delta^3\sqrt{\phi^2-\theta^2}}\,.
\end{equation}
The obtained expression demonstrates the fact that this difference specifying the thickness of the region where the
diffusion flux threshold exhibits complex behavior is extremely narrow (Fig.~\ref{Fig.5}). It is of the third order in
the small parameter $1/\Delta$ and can be ignored at all. In this case only the first term in expansion~\eqref{ib:72}
should be taken into account, thus, only branch~\eqref{ib:73} exists. So by virtue of \eqref{ib:77} the diffusion flux
threshold in the layer $\mathcal{L}_\phi$ as well as in its small neighborhood meeting the interval $\sigma_m < \sigma
\ll 1$ is approximated by the expression
\begin{equation}\label{ib:81}
    g_{c}(\sigma) \approx \frac{16\theta_{c0}^3}{\sqrt{\sigma^2-\sigma_m^2}\,\left(\sigma +
    \sqrt{\sigma^2-\sigma_m^2}\right)^2}
\end{equation}
showing some formal singularity when $\sigma\to \sigma_m+0$.

Finalizing this subsection let us discuss the behavior of the instability boundary $\mathcal{B}_\phi$ depending on the parameter $\Delta$ including its relatively small values. It should be reminded that previously we considered two curves on the plane $\{\theta,\phi\}$, the instability boundary $\mathcal{B}_\phi = \{\phi_c(\theta)\}$ itself and the curve $\mathcal{B}_\phi^+ = \{\phi^+_c(\theta)\}$. The former singles out the points on this plane where the crystal growth can become unstable for some values of the species diffusion flux. The latter is the boundary of the growth instability for vary large values of the diffusion flux, $g\to\infty$.

\begin{figure}
\begin{center}
\includegraphics[width=60mm]{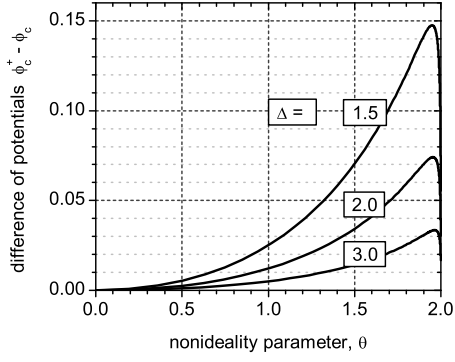}
\end{center}
\caption{The difference in the instability boundaries $\mathcal{B}^+_\phi$ and $\mathcal{B}_\phi$ for several values of the parameter $\Delta$.}
\label{Fig.14}
\end{figure}

It can be demonstrated analyzing directly the general eigenvalue equation~\eqref{ib:7} and the expression~\eqref{ib:8} for the critical diffusion flux $g_c$ that the terminal points of the curves $\mathcal{B}_\phi$ and $\mathcal{B}_\phi^+$ at $\theta=0$ and  $\theta=\theta_{c0}$ coincide with each other for a given value of $\Delta$. So by virtue of expression~\eqref{verynew:2} their coordinates are specified by the following expressions
\begin{align}
\label{new:6a}
    \phi_{0} & = 2\theta_{c0} \frac{\Delta^2+ \sqrt2\Delta+1}{\Delta^2-1}\, && \text{for $\theta = 0$}
\\
\label{new:6b}
        \phi_{c0} & = \theta_{c0}\frac{\Delta^2 +1}{\Delta^2 -1}\, &&\text{for $\theta = \theta_{c0}$}\,.
\end{align}
As it must the $\phi$-coordinates of both the points have a singularity as $\Delta\to1$ because in this limit the growth is stable for $\theta<\theta_{c0}$.

For the intermediate points $0<\theta<\theta_{c0}$ the curves $\mathcal{B}_\phi$, $\mathcal{B}_\phi^+$ deviate from each other. To evaluate this difference Fig.~\ref{Fig.14} plots the difference $\phi^+_{c}-\phi_c$ vs the potential $\theta$ for several value of $\Delta$. As seen in Fig.~\ref{Fig.14} the curves $\mathcal{B}_\phi$ and $\mathcal{B}^+_\phi$ practically coincide with each other except for the values of $\Delta$ coming too close to its threshold $\Delta=1$. Thereby expression~\eqref{verynew:2} gives a fairly fine approximation of the instability boundary $\mathcal{B}_\phi$ for such values of $\Delta$.

\subsection{Double critical point and its neighborhood $\mathcal{C}$}

\begin{figure}
\begin{center}
\includegraphics[width=60mm]{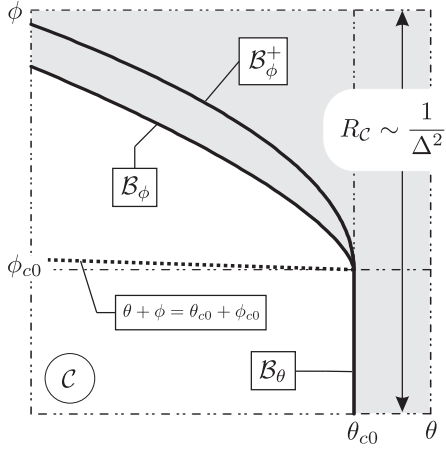}
\end{center}
\caption{Structure of the instability region in a close proximity to the double critical point
$\{\theta_{c0},\phi_{c0}\}$. }
\label{Fig.7d}
\end{figure}

The boundaries $\mathcal{B}_\theta$ and $\mathcal{B}_\phi$ of the instability region meet at the point
$\{\theta_{c0},\phi_{c0}\}$ that can be referred to as a double critical point because its coordinates are the
threshold of the nonideality parameter and the threshold of the asymmetry potential exceeding which the system
asymmetry changes the instability property substantially. The latter implies the fact that the asymmetry causes the
instability onset in the system being stable before the potential exceeds the threshold, $\phi > \phi_{c0}$, and
$\phi_{c0}$ is the minimal value possessing this property among all the possible values of the solid composition $\chi$
and the nonideality parameter $\theta$. Therefore in calculating the value of $\phi_{c0}$ we can set
$\theta=\theta_{c0}$.

The critical region $\mathcal{C}$ is a certain neighborhood of the point $\{\theta_{c0},\phi_{c0}\}$ where the layers
$\mathcal{L}_\theta$ and $\mathcal{L}_\phi$ overlap with each other. So it should exhibit some crossover between the
properties of these layers. According to the results to be obtained in the region $\mathcal{C}$ the potential
difference $\phi-\theta$ is rather small so not only the inequality  $\zeta_c\Delta\gg1$ but also
$\zeta_c/\Delta\gg1$. Keeping in mind the general condition~\eqref{ib:11} necessary for the system asymmetry to
affect essentially the instability onset we describe the region $\mathcal{C}$ with two small parameters $u\ll1$ and
$v\ll1$ introduces as follows
\begin{equation}\label{ib:100}
\begin{split}
    (\phi-\theta) & = \frac{\phi+\theta}{\Delta^2}(1+u)\,,
\\
    (\theta-\theta_{c0}) & = \frac{\phi+\theta}{2\sqrt2\Delta^2}v\,.
\end{split}
\end{equation}
Then for the variable $\xi:= \Delta/\zeta \ll1$ regarded as a small value the eigenvalue equation~\eqref{ib:7} is
reduced to
\begin{equation}\label{ib:101}
   v = -x u + x^3
\end{equation}
and expression~\eqref{ib:8} for the diffusion flux threshold takes the form
\begin{equation}\label{ib:102}
    g_{c\{\mathcal{C}\}} = \frac{4\sqrt2\Delta^4}{(\theta+\phi)}\,\frac1{x^2(x\sqrt2-u)}\,.
\end{equation}
As it must, when $u<0$ the instability boundary is specified by the equality $v=0$ ($\theta = \theta_{c0}$) and the
diffusion flux threshold $g_c\to\infty$ as $v\to+0$. For $u>0$ the system changes the behavior.

The eigenvalue equation~\eqref{ib:101} relating the variables $u$ and $v$ at the point $x = \sqrt{u/3}$ where its
right-hand side attains the minimum specifies the instability boundary $\mathcal{B}_\phi$, namely,
\begin{subequations}\label{ib:103}
\begin{align}
    \label{ib:103a}
    v & = -\frac{2}{3\sqrt3} u^{3/2}
\\
\intertext{or returning to the variables $\theta$ and $\phi$}
    \label{ib:103b}
    \frac{\theta-\theta_{c0}}{\theta_{c0}}& = - \frac{\Delta}{6\sqrt3}
    \left( \frac{\phi -\phi_{c0}}{\phi_{c0}} \right)^{3/2}\,.
\end{align}
\end{subequations}
As should be expected, at the boundary $\mathcal{B}_\phi$ the diffusion flux threshold takes a finite value equal to
\begin{equation}\label{ib:104}
    g_{c\{\mathcal{C}|\mathcal{B}_\phi\}} = 6\sqrt6\Delta \left(\frac{\phi_{c0}}{\phi - \phi_{c0}}\right)^{3/2}\,.
\end{equation}
Naturally, the diffusion flux threshold $g_c$ diverges as the asymmetry potential $\phi\to\phi_{c0}+0$.

Near the boundary $\mathcal{B}_\phi$ the values of the diffusion flux causing the instability onset are bounded from
below and above. The locus $\mathcal{B}_\phi^+$ where the upper boundary goes to infinity is specified by the
singularity point of function~\eqref{ib:102}, i.e. $x= u/\sqrt2$. This value via equality~\eqref{ib:101} gives us the
relationship between the potentials $\theta$ and $\phi$ at the curve $\mathcal{B}_\phi^+$
\begin{subequations}\label{ib:105}
\begin{align}
\label{ib:105a}
       v &=- \frac1{\sqrt2} u^2
\\
\intertext{or}
\label{ib:105b}
    \frac{\theta-\theta_{c0}}{\theta_{c0}}& = - \frac{\Delta^2}{16}
    \left( \frac{\phi -\phi_{c0}}{\phi_{c0}} \right)^2\,.
\end{align}
\end{subequations}
The expressions obtained here hold for $u,v\ll1$, so the characteristic size of the region of double criticality is
about $R_c\sim 1/\Delta^2$.

\section{Regimes of instability dynamics}\label{Sec:scenarios}

The present section is devoted to a qualitative analysis of the system dynamics. For the sake
of simplicity we ignore difference in the species kinetic coefficients setting $D_1 = D_2 = D$ and $\nu_1=\nu_2$.

\begin{figure}
\begin{center}
\includegraphics{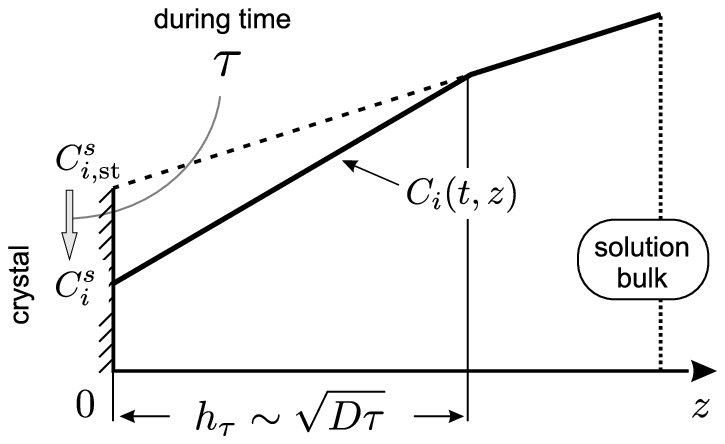}
\end{center}
\caption{Perturbation of the species distribution in the aqueous solution induced by variations in the surface
concentration $C^s_i$ on time scales about $\tau$. Schematic illustration.}
\label{Fig.2}
\end{figure}

At first, let us consider perturbations of the species distribution $\delta C_i(z,t)$ induced by small variations $\delta\chi(t)$ in the surface composition on time scales about $\tau$. Actually $1/\tau$ is the perturbation increment analyzed in the previous Section. Change in the
surface composition $\chi(t)$ affects directly the species attachment rate caused by the growth process, which, in
turn, gives rise to variations in the species concentration near the crystal surface $\delta C_i^s$. These boundary
variations in the species concentration spread into the solution bulk, which is responsible for the formation of
spatial perturbations in the species distribution schematically shown in Fig.~\ref{Fig.2}. The characteristic spatial
scale of these perturbations can be estimated as $h_\tau\sim (D\tau)^{1/2}$.

Within a qualitative approximation mass conservation for such perturbations reads
\begin{equation}\label{spec:0}
    -\frac{h_\tau \delta C_i^s}{\tau}\sim \delta\left[ r_i(C_i^s,\chi)\right]
\end{equation}
or, by virtue of \eqref{attrate},
\begin{equation}\label{spec:1}
    -\frac{h_\tau}{\tau}\delta C_i^s\sim
    \frac{a}{\tau_i}\delta C_i^s + \frac{aC_{i,\text{st}}}{\tau_i}\omega_i\delta\chi\,,
\end{equation}
where the quantities (for $i=1$,~2)
\begin{equation}\label{spec:2}
    \omega_i(\chi_{\text{st}}) = -\left.\frac{d\ln \tau_i(\chi)}{d \chi}
    \right|_{\chi=\chi_\text{st}}
\end{equation}
have been introduced and by virtue of \eqref{tau}
\begin{align}\label{spec:3}
    \omega_1 & = \phi + \theta\,, & \omega_2 & = \phi - \theta \,.
\end{align}
Expression~\eqref{spec:1} enables us to single out two limit cases. The first one which will be referred to as
\emph{the growth regime of constant growth rate} matches rather slow variations of the crystal composition $\chi$ and
the species concentration $C_i$, namely, the condition $\tau\gg \tau_i(h_\tau/a)$ or, what is the same,
\begin{equation}\label{lcg:1}
    \tau \gg \frac{D\tau_i^2}{a^2}\,.
\end{equation}
In this case \eqref{spec:1} yields
\begin{gather}
    \label{spec:4a}
        \delta C_i^s \approx - C_{i,\text{st}}^s \omega_i \delta \chi
\\
\intertext{and, thus, via \eqref{spec:0}}
        \delta [r_i(C_i^s,\chi)] \sim  \left(\frac{D\tau_i^2}{\tau a^2}\right)^{1/2}\cdot  r_{i,\text{st}}
        \ll r_{i,\text{st}}\,.
    \label{spec:4b}
\end{gather}
Thereby for slow variations of the crystal composition $\chi$ the induced perturbations in the species distribution $C_i(z,t)$ are in quasiequilibrium.  In other words, the boundary value $C_i^s$ of the species concentration changes in time with $\chi$ in such a manner that the boundary value of the diffusion flux, the species attachment rate $r_i$, be practically equal to the inflow of the corresponding species at distant points. In particular, exactly such variations are described by expression~\eqref{spec:4a} being linearization of the condition
\begin{equation}\label{spec:4c}
    \frac{aC^s_i}{\tau_i(\chi)} = r_i\approx\text{const}\,.
\end{equation}
The second limit case, which will be called \emph{the growth regime of constant surface concentration} is related to
rather fast variations in the crystal composition $\chi$, when their time scale $\tau$ meets the inequality $\tau \ll
\tau_i(h_\tau/a)$ or
\begin{equation}\label{lcg:2}
    \tau \ll \frac{D\tau_i^2}{a^2}\,.
\end{equation}
In this case the induced variations in the surface concentration $C^s_i$ of species $i$ are rather small in comparison
with that could be expected in the previous limit case,
\begin{equation}\label{lcg:3}
   \delta C_i^s \sim -\left(\frac{\tau a^2}{D\tau_i^2}\right)^{1/2}\cdot\omega_i C^s_{i,\text{st}}\delta\chi
   \ll \omega_i C^s_{i,\text{st}}\delta\chi\,.
\end{equation}
Therefore, at the first approximation the fast dynamics of the species distribution and the crystal composition meets
the equalities
\begin{align}\label{lcg:4}
    C^s_i &\approx \text{const} & &\text{and} &\delta r_i &
    \approx aC^s_i\delta \left[\frac1{\tau_i(\chi)}\right]\,.
\end{align}

In this consideration the variations of the crystal composition $\chi(t)$ were treated to be given beforehand. In order to draw some conclusions about the growth dynamics as a behavior of an autonomous system it is necessary to discuss how the induced variations of the attachment rates $r_1$ and $r_2$ affect, in their turn, the crystal composition $\chi$. This effect is described by the governing equation~\eqref{chidot}.

\begin{figure}
\begin{center}
\includegraphics[width = 0.8\columnwidth]{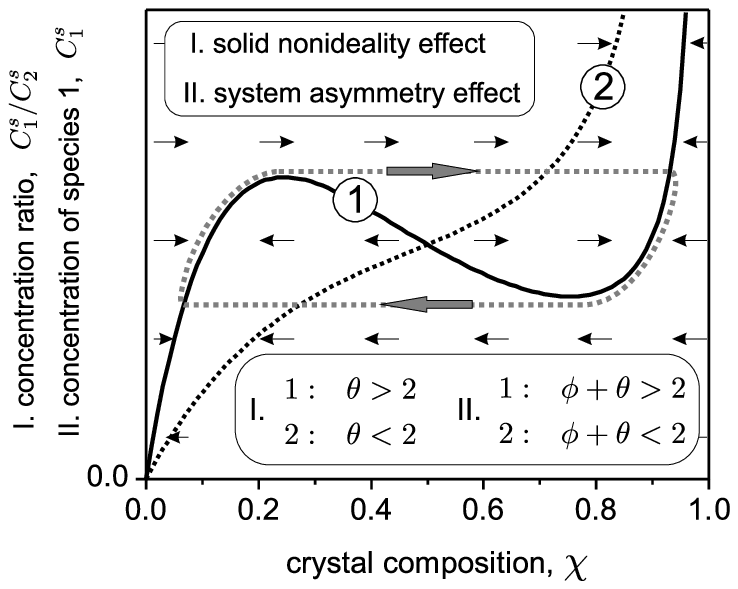}
\end{center}
\caption{Phase planes demonstrating the mechanism of the instability onset caused by the solid nonideality (I) and the
system asymmetry (II).}
\label{Fig.3}
\end{figure}

As was discussed in the previous Section, when the nonideality potential exceeds the critical value, $\theta>\theta_{c0}$ the perturbation increment $1/\tau\to \infty$ as the species diffusion flux goes to infinity also. So it is natural to expect that for the developed instability the regime of constant surface concentration takes place with respect to \emph{both} the species components. Then keeping in mind expressions~\eqref{lcg:4} and applying to equation~\eqref{chidot} governing the dynamics of crystal composition we can draw the velocity field of the system motion on the phase plane $\{C_1^s/C_2^s, \chi\}$ as shown in Fig.~\ref{Fig.3}. The curve
\begin{equation}\label{qualtheta:1}
    \frac{C_1^s}{C_2^s} = \frac{\chi}{(1-\chi)}\,\frac{\tau_1(\chi)}{\tau_2(\chi)}
    \propto \frac{\chi}{(1-\chi)} e^{-2\theta\chi}
\end{equation}
divides this phase plane into parts with the opposite directions of the velocity field. In obtaining \eqref{qualtheta:1} expressions~\eqref{tau} have been used. As it should be the stationary values of the species concentrations $C_{i,\text{st}}^s$ and the crystal composition $\chi_\text{st}$ (see expressions~\eqref{spec:4c}) meet equality~\eqref{qualtheta:1}. Figure~\ref{Fig.3} clearly demonstrates us that under such conditions its increasing branches are stable whereas a decreasing branch (if it exists) is unstable. So the limit circle at a rough approximation should have the form shown in Fig.~\ref{Fig.3}. Exactly this limit was analyzed in our previous paper~\cite{we} and corresponds to the domain $\mathcal{D}_\theta$ of the instability region.

If the nonideality parameter $\theta$ is less then the critical value, $\theta < \theta_{c0} = 2$ the solid nonideality cannot itself induce the growth instability. In this case the instability development is governed by the system asymmetry, which is reflects in properties of the instability domain $\mathcal{D}_\text{0--1}$. In particular, for the system with such parameters only the channel 0--1 of the precipitation reactions~\eqref{Intr:5} plays an active role, the channel 0--2 is characterized by the equilibrium value of the species diffusion flux at the crystal surface. In this case it is quite natural to assume that the perturbation increment $1/\tau$ meets the inequality
\begin{equation}\label{assym:1}
    \frac{D\tau_2}{a^2} \ll \tau \ll \frac{D\tau_1}{a^2}\,.
\end{equation}
Therefore, on one hand, with respect to species 2 such a process can be classified within the regime of constant growth rate. On the other hand, with respect to species 1 the regime of constant surface concentration takes place. Actually it is the case for the points of the domain $\mathcal{D}_\text{0--1}$. To describe the corresponding dynamics of the crystal composition $\chi$ we can fix the surface concentration $C_1^s$ and set the species attachment rate $r_2 = (1-\chi_\text{st})G$. Then we draw a similar velocity field of the system motion on the phase space $\{C_1^s,\chi\}$ shown again in the same Fig.~\ref{Fig.3}. Its pattern is identical to one discussed above except for the fact that the $y$-axis of this phase plane has now another meaning, it presents the surface concentration of species 1. As follows from equation~\eqref{chidot} and expressions~\eqref{tau} the curve
\begin{equation}\label{assym:2}
    C_1^s = \frac{\chi}{(1-\chi)}\,\tau_1(\chi)\chi_\text{st}G
    \propto \frac{\chi}{(1-\chi)} e^{-(\theta +\phi)\chi}
\end{equation}
separates the regions on the phase plane $\{C_1^s,\chi\}$ with the opposite directions of the velocity field. This
curve looks like the previous one~\eqref{qualtheta:1} within the replacement $2\theta \rightarrow\theta+\phi$. So again the instability condition for the potentials of the species interactions take the form $\theta+\phi > 2\theta_{c0}$, being in agreement with the results obtained before. As previously the increasing branches of
curve~\eqref{assym:2} are stable whereas the decreasing one is unstable and the system transition between them as well
as the transition from the unstable stationary point $\{C^s_{1,\text{st}},\chi_\text{st}\}$ to one of them proceeds
within the regime of constant surface concentration with respect to species 1. The rough approximation of the limit circle again has the same form.

In the part of the domain $\mathcal{D}_\text{0--1}$ where $\theta>\theta_{c0}$ both of the instability scenarios can be implemented. So depending on the species diffusion flux either the phase plane $\{C_1^s,\chi\}$ or the plane $\{C_1^s/C_2^s, \chi\}$ can give an appropriate representation of the system dynamics.

\section{Nonlinear dynamics of system instability. Domain $\mathcal{D}_\phi$}

This section presents numerical results for the system dynamics when the growth instability arise in a subdomain
$\mathcal{D}_\phi$ of the domain $\mathcal{D}_\text{0-1}$, where the nonideality parameter $\theta$ is less then its
threshold, i.e. $\theta < \theta_{c0}$. So it is the the system asymmetry that causes the instability.

To model numerically the system dynamics the governing equations~\eqref{Diffusion}--\eqref{boundaryfin}, and
\eqref{chidot} were converted into dimensionless form. Namely, first, the time $t$ and the spatial coordinate $z$ are
measured in units
\begin{align}\label{nd:1}
   \tau^* & = \frac{\sqrt{D_1D_2}\tau^2_g}{a^2}\,,& z^* & =\frac{\sqrt{D_1D_2}\tau_g}{a}\,,
\end{align}
respectively, i.e. the dimensionless time and spatial coordinates are introduced as $t_\text{new} = t_\text{old}/\tau^*$ and $z_\text{new}=z_\text{old}/z^*$.
Second, the species concentrations and the diffusion flux are replaces with their dimensionless analogies, $C_{i,\text{new}} =
C_{i,\text{old}}/C^*$ and $G_{i,\text{new}}=G_{i,\text{old}}/G^*$, where
\begin{align}\label{nd:2}
    C^* & = \frac{1}{\sqrt{D_1D_2}\tau_g a}\,, &     G^* & = \frac{1}{\sqrt{D_1D_2}\tau^2_g} \,.
\end{align}
In this way the original model is rewritten in the form
\begin{align}
   \label{nd:3a}
     \frac{\partial C_i}{\partial t} & = \kappa_i\frac{\partial^2 C_i}{\partial z^2}\,,
\\
   \label{nd:3b}
     \frac{d\chi}{dt} & =
     \Bigl[
     \left(1-\chi\right)\varrho_1(\chi)\,C^s_1-\chi\varrho_2(\chi)\,C^s_2
     \Bigr]\,,
\end{align}
with equation~\eqref{nd:3a} being subject to the boundary condition at $z=0$
\begin{equation}\label{nd:4}
    \kappa_i\left. \frac{\partial C_i}{\partial z}
    \right|_{z=0} = \varrho_i(\chi) C^\mathrm{s}_i
\end{equation}
and the condition at distant points, i.e. at the formal external boundary $L_\text{new} = L_\text{old}/z^*$,
\begin{equation}\label{nd:5}
G_i = \kappa_i\left. \frac{\partial C_i}{\partial z} \right|_{z=L}\,.
\end{equation}
Here the dimensionless species diffusivities are
\begin{gather}
    \label{nd:6}
    \kappa_1 = \frac1{\kappa_2} = \sqrt{\frac{D_1}{D_2}}
\\
\intertext{and the dimensionless rates of the atom attachment to the growing crystal are}
    \label{nd:7}
\begin{split}
    \varrho_1(\chi) & = \varkappa_1 \exp\left\{ \phi\chi - \theta(1-\chi)\right\}\,,
\\
    \varrho_2(\chi) & = \varkappa_2 \exp\left\{ -\phi(1-\chi) - \theta\chi)\right\}
\end{split}
\\
\intertext{with}
\label{nd:8}
     \varkappa_1 = \frac1{\varkappa_2} =
     \left(\frac{\nu_1}{\nu_2}\right)^{1/2} \exp\Big\{-\frac12\eta\Big\}\,.
\end{gather}
It should be noted that the previously used parameter $\Delta_\phi$ is related to the introduced kinetic coefficients
as
\begin{equation}\label{nd:9}
    \frac{\varkappa_2}{\varkappa_1}= \left(\frac{\kappa_2}{\kappa_1}\right)^{1/2} e^{\phi}\Delta^2_\phi\,.
\end{equation}
So the ratio $(\varkappa_1/\varkappa_2)e^\phi$ is actually the main small parameter of the given model because for aqueous
solutions the relationship $D_1\sim D_2$ is typically fulfilled.

The system of equations~\eqref{nd:3a}--\eqref{nd:5} was solved numerically using the Crank-Nicholson scheme for the
diffusion equation~\eqref{nd:3a} and the midpoint method for equation~\eqref{nd:3b}. To exemplify the basic
characteristics of the instability dynamics in the region $\mathcal{D}_\phi$ the system parameter were set equal to
$\theta = 1.5$, $\phi = 3.5$, and $\Delta_\phi = 10$ as well as $\kappa_1=\kappa_2 = 1$. Then expression~\eqref{nd:9}
gave us the values of $\varkappa_1$ and $\varkappa_2$. The time and spatial steps in the simulation routine were 0.01,
decreasing the steps twice did not affect the obtained results. The time variations in the species distribution induced
by the developed instability turned out to be located near the crystal boundary within a layer of thickness about
15--20~spatial units. So the external boundary of the system was placed at $L=100$, where the species concentrations
$C_1^\infty$ and $C_2^\infty$ were fixed in such a way that the total diffusion flux and the solid composition take the
values $G^\text{st}= G^\text{st}_1+G^\text{st}_2 = 10$ and $\chi^\text{st} = 0.5$ under the steady state conditions.
The total simulation time was 10000 time units.

\begin{figure}[t]
\begin{center}
\includegraphics[width=\columnwidth]{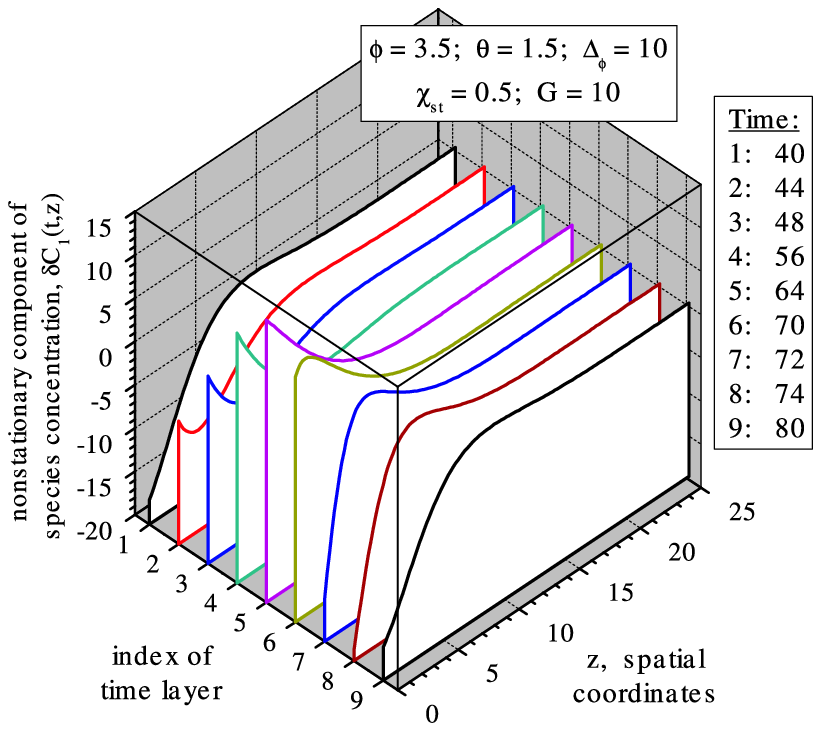}
\end{center}
\caption{The time dependent component $\delta C_1(t,z)$ of the species 1 distribution in the aqueous solution bulk near
the crystal surface, $z=0$, for several time slices within one period of the oscillations. Result of numerical
simulation. The shown time origin $t=0$ is placed at an arbitrary chosen point that corresponds to the instability
becoming well developed.}
\label{Fig.11}
\end{figure}

\begin{figure*}
\begin{center}
\includegraphics[width=0.8\textwidth]{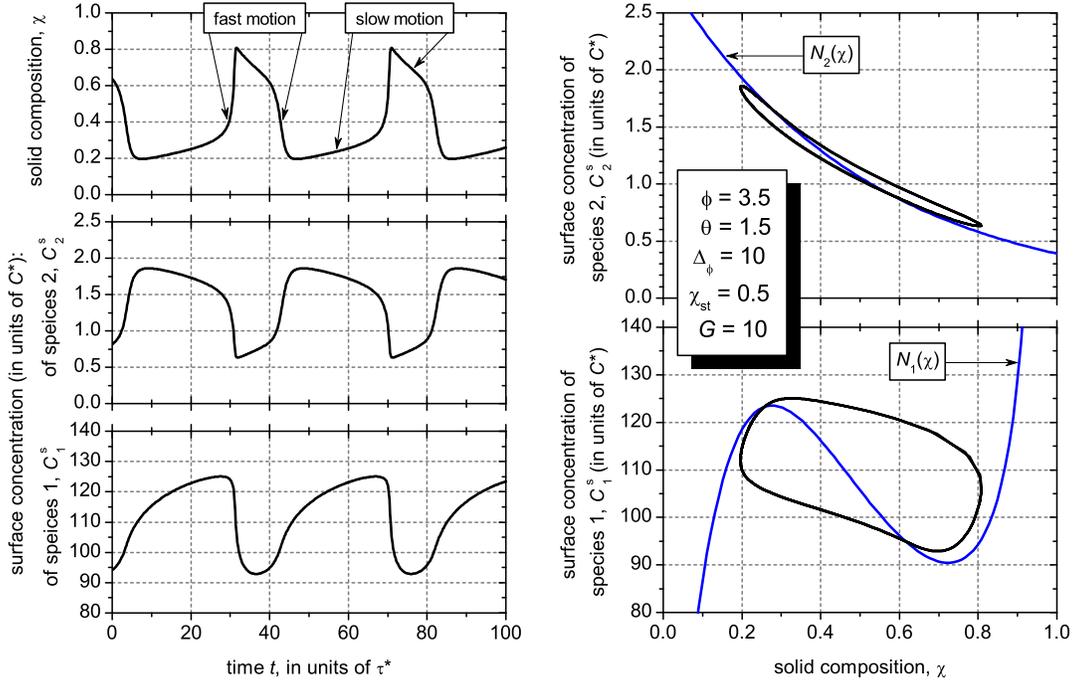}
\end{center}
\caption{The dynamics of solid composition $\chi(t)$ and the surface species concentrations $C_i^s(t)$ (left column)
and the corresponding phase portraits on the planes $\{\chi,C_i^s\}$ (right column). Result of numerical simulation.
The shown time origin $t=0$ is placed at an arbitrary chosen point that corresponds to the instability becoming well
developed and steady state.}
\label{Fig.10}
\end{figure*}

Below we will present the obtained results. Figure~\ref{Fig.11} visualizes evolution of the species distribution in the aqueous solution bulk near the crystal surface. Only the distribution of species 1 is shown because it, first, exemplifies similar effects for species 2 also and, second, plays the leading role in the instability onset. To elucidate the dynamics of the species distribution the time dependent component $\delta C_1(t,z)$ is singled out from the total distribution function
\begin{equation*}
    C_1(t,z) = \delta C_1(t,z) + \left<C_1^s\right> + \frac{\left<G_1\right>}{\kappa_1}z
\end{equation*}
and depicted in Fig.~\ref{Fig.11}. The other terms in this expression are the steady state components of the species distribution. As seen in this figure the time variations of species distribution are located near the crystal surface $z=0$ in its neighborhood of thickness about $L_C\sim 15$ for the chosen system parameters. So the size of the system $L=100$ used in the numerical simulations is fairly large to enable one to regard the external boundary $z=L$ as infinitely distant points. In any case in numerical simulations the size of the system should be specified that the inequality $L_C\lesssim L$ to hold.

Figure~\ref{Fig.11} demonstrates us the fact that a simple model of the boundary layer similar to the one shown in Fig.~\ref{Fig.2} can be used only for a qualitative analysis. The actual spatial form of $\delta C_1(t,z)$ can possess a remarkable extremum attained at a certain internal point of the crystal neighborhood, which must be taken into account in constructing an appropriate boundary layer approximation.

Nevertheless, in spite of a rather rough model for the boundary layer used in Sec.~\ref{Sec:scenarios} the instability
scenarios described there is justified by the results of numerical simulation. The found dynamics of the solid
composition $\chi(t)$ and the surface species concentrations $C^s_i(t)$ exhibit relaxation oscillations with clearly
visible fast and slow stages of system motion (Fig.~\ref{Fig.10}). So the results obtained for the given set of
parameters do describe an essentially nonlinear regime of the growth instability. The phase portrait of the system
oscillations on the plane $\{\chi,C_2^s\}$ demonstrates us the fact that the regime of constant growth rate really
takes place with respect to the species 2. Indeed the image of the oscillation limit circle on this phase plane is
located in the vicinity of the curve $N_2(\chi)$ obtained by setting the right-hand side of the boundary
condition~\eqref{nd:4} equal to the diffusion flux of species~2 under the stationary conditions, i.e.
\begin{align}
\nonumber
     \varrho_2(\chi)C_2^s & = (1-\chi_\text{st})G
\\
\intertext{and thus}
\label{nd:10}
     N_2(\chi) & = \frac{(1-\chi_\text{st})G\,e^\phi}{\varkappa_2}\cdot\exp\left\{- (\phi-\theta)\chi\right\}\,.
\end{align}

With respect to species 1 the regime of constant surface concentration could be expected to be the case. The image of
the oscillation limit circle on the plane $\{\chi,C_1^s\}$ (Fig.~\ref{Fig.10}) justifies this expectation at least
within semi-quantitative consideration. Figure~\ref{Fig.10} depicts the obtained limit circle together with the
nullcline $N_1(\chi)$ constructed by setting the right-hand side of the governing equation~\eqref{nd:3b} equal to zero,
fixing the surface concentration $C_1^s$ and assuming the attachment rate $\varrho_2(\chi)C_2^s$ of species 2 to meet
the regime of constant diffusion flux. In this the expression
\begin{equation}\label{nd:11}
    N_1(\chi) = \frac{(1-\chi_\text{st})G\,e^\theta}{\varkappa_1}\cdot \frac{\chi\exp\left\{ - (\phi +\theta)\chi
    \right\}}{(1-\chi)}
\end{equation}
has been constructed. As seen, here the fragments of the limit circle matching the fast motion deviate substantially
from the decreasing branch of the nullcline $N_1(\chi)$ and the fragments of slow motion go near its increasing
branches. So, roughly speaking, it is the characteristics of the nullcline $N_1(\chi)$ that specify the amplitudes of
time variations in the solid composition and surface species concentrations for the developed growth instability.
However, the obtained limit circle also deviates remarkably from a simple form constructed in Fig.~\ref{Fig.3} applying
directly to the notions of the standard relaxation oscillations. The matter is that the system under consideration is
really not reduced to a two-variable model implying actually the too simple boundary layer approximation shown in
Fig.~\ref{Fig.2} to hold. So the dynamics of the surface concentration $C^s_1(t)$ of species~1 contains the fragments
of slow motion as well as that of fast motion (Fig.~\ref{Fig.10}). The latter ones actually force the fast motion
branches of the limit circle to deviate remarkable from horizontal lines on the plane $\{\chi,C_1^s\}$. This effect was
also observed for the growth instability caused by the solid nonideality \cite{we}.

\begin{figure}
\begin{center}
\includegraphics[width=0.8\columnwidth]{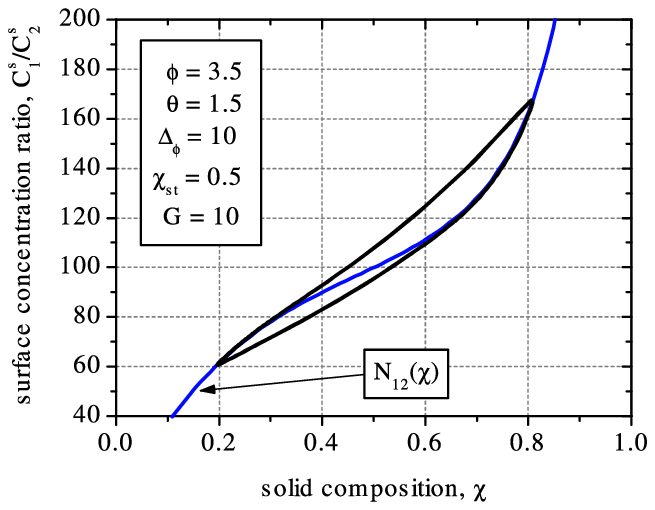}
\end{center}
\caption{The phase portrait of the system oscillations on the plane $\{\chi,(C^s_1/C_2^s)\}$. Results of numerical
simulation.}
\label{Fig.12}
\end{figure}

Finalizing the present Section we underline once more that there is a widely used approach to constructing the limit
circle of oscillations in such system, i.e. the ``boundary reaction -- diffusion'' systems treating the governing
equation~\eqref{nd:3b} (or its original version~\eqref{chidot}) for the solid composition in a too simple way. It sets
the right-hand side of this equation equal to zero and relates the system portrait on the plane
$\{\chi,(C^s_1/C_2^s)\}$ to the nullcline $N_{12}(\chi)$ determined by the expression
\begin{multline}\label{nd:12}
    N_{12}(\chi) = \frac{\chi}{(1-\chi)}\cdot\frac{\varrho_2(\chi)}{\varrho_1(\chi)}
\\
    = \frac{\varkappa_2e^{-\phi}}{\varkappa_1}\cdot \frac{\chi\exp\{\theta(1-2\chi) \}}{(1-\chi)}\,.
\end{multline}
For the growth instability caused the solid nonideality the nullcline $N_{12}(\chi)$ possesses a decreasing branch being unstable (see, e.g. Fig.~\ref{Fig.3}). In this case the limit circle constructed following the classical ideas of the standard relaxation oscillations is justified at least within a quasi-qualitative analysis \cite{we}. However, if the growth instability is induced by the system asymmetry, such an approach is not justified at all, the corresponding nullcline $N_{12}(\chi)$ is a monotonous curve and the system portrait on the plane $\{\chi,(C^s_1/C_2^s)\}$ is just located in its vicinity (Fig.~\ref{Fig.12}).

\section{Conclusion}

We have analyzed the oscillatory zoning, i.e. the self-organization phenomenon arising during crystallization of multi-component solid from aqueous solution. It manifests itself in self-formation of quasi-periodic spatial pattern of solid composition from the core of a crystallite to its rim.

Keeping in mind systems like (Ba,Sr)SO$_4$ we have proposed a model for the growth of ternary-component solid from aqueous solution. The crystallization process comprises passive diffusion of species towards the crystal surface through the aqueous solution bulk, their adsorption at the crystal surface, and incorporation into the crystalline lattice at the surface atomic steps. The latter process is assumed to limit the crystal growth, so the species adsorption-desorption at the crystal surface is described within the quasi-equilibrium approximation. Due to a very low rate of crystallization from aqueous solutions the growth dynamics is simulated using the boundary-reaction-diffusion model for the species distribution in the aqueous solution bulk.

The proposed model for the growth process takes into account the solid nonideality as well as the system asymmetry, with the latter being the characteristic feature of systems for which oscillatory zoning was reproduced in laboratory also. It has been demonstrated that the system asymmetry can cause the growth instability in the case when the solid nonideality is low, i.e. the nonideality parameter is less than its threshold, $\theta<\theta_{c0}$, or even if the solid solution is ideal, $\theta = 0$. Using the linear stability analysis the instability domain is constructed in the phase space $\{\theta,\phi,\Delta,g\}$ comprising the nonideality parameter $\theta$, the difference $\phi$ of the species interaction constants, the parameter $\Delta$ characterizing the ratio between time scales of species incorporation into the crystalline lattice, and the species diffusion flux (in dimensionless units). The potential difference $\phi>0$ is assumed beforehand to be nonnegative because, otherwise, exchanging the species indices makes it value positive.  Projection of this domain onto the plane $\{\theta,\phi\}$ for a fixed value of $\Delta$ enables us to divide all the points on the plane $\{\theta,\phi\}$ into stable and unstable ones. The latter points correspond to such solids for which the growth instability under consideration can arise in principle.

It has been demonstrated that there are five characteristic regions on the plane $\{\theta,\phi\}$, where the growth instability exhibits different properties. In particular, in the region
$$
    \{\theta>\theta_c;\,\theta+\phi < 2\theta_{c0}\}
$$
the growth instability is governed mainly by the solid nonideality and was analyzed in detail previously in Ref.~\cite{we}. In the region
$$
    \{\theta>\theta_c;\,\theta+\phi > 2\theta_{c0}\}
$$
for $\Delta \gg 1$ the instability onset is governed by the system asymmetry and, as a result, only one species plays an active role, the diffusion flux of the other component is practically quasiequilibrium. However for large values of the diffusion flux the instability dynamics again is mainly affected by the solid nonideality. In the region
$$
    \{\theta<\theta_c;\,\theta+\phi > 2\theta_{c0}\}
$$
for $\Delta\gg1$ the instability is due to the system asymmetry even for large values of the diffusion flux. It can arise also for the ideal solid solution. In this case the the critical value $g_c$ of the species diffusion flux exhibits a rather complex behavior neat the instability boundary, in particular, $g_c$ remains bounded as the system comes close to it. It has demonstrated that the system asymmetry can induce, in principle, the growth instability if $\Delta >1$, however if $\Delta\to1$ the required value of the potential difference $\phi_{c}\to\infty$ (for a fixed value of $\theta < \theta_{c0}$). The condition that the system admits an unstable perturbation with finite spatial scales for large values of the species diffusion flux, $g\to\infty$, gives a fairly precise approximation of the boundary of the instability caused by the system asymmetry except for values of $\Delta$ close to its threshold $\Delta =1$.

Analyzing the limits cases of the growth dynamics two typical regimes were singled out. One of them is the regime of constant diffusion flux that characterizes ``slow'' dynamics of species concentration and solid composition. The other referred to as the regime of constants surface concentration described the stage of ``fast'' dynamics. Oscillatory zoning studied in our previous paper~\cite{we} corresponds to the case when the region of constant surface concentration holds with respect to all the species. As a result the phase portrait of the system dynamics looks line a limit circle of relaxation oscillations on the phase plane $\{C_1^s/C_2^s,\chi\}$. At a rough approximation it can be constructed referring to the $N$-like curve showing the quasi-stationary dependence of the ratio $C_1^s/C_2^s$ on $\chi$.  In the present paper the main attention is paid to the case $\Delta\gg1$ where the nonlinear stage of the developed instability is characterized by the regime of constant surface concentration with respect to one species and regime of constant diffusion flux with respect to the other species. Now the phase plane $\{C_1^s,\chi\}$ gives the appropriate representation of the system portrait in a similar way, including the construction of the limit circle describing oscillatory zoning.

Numerical simulation justifies these conclusions. Besides, the species distribution in the aqueous solution bulk found numerically demonstrates the fact that a rather sophisticated model of the boundary layer should be developed to describe oscillatory zoning adequately.

At the next of the theory development OZ in 2D case will be considered with respect to two aspects. One is the affect of 2D species distribution itself on the pattern formation. The other is due to the fact that, for example, in the domain $\mathcal{D}_\text{0--1}$ the ``optimal'' conditions for the instability development can match the solid composition $\chi\ne0.5$. In this case a special instability with respect to nonuniform perturbations along the crystal surface can arise. So we expect that the characteristic length determining spatial correlations of the OZ-pattern will be found in this way. Nevertheless referring to OZ obtained in laboratory \cite{Putn1,Putn2,Putn3} this length should be expected to exceed essentially the size of growing crystallites about 200~$\mu$m.

\acknowledgments

One of the authors (IL) appreciates the financial support of the SFB 458 and the University of Munster as well as the partial support of DFG Grant MA 1508/8-1 and RFBR Grant 06-08-89501. The authors also thank Putnis for helpful discussions.

\end{document}